\documentclass[aip,graphicx]{revtex4-1}

\draft 
\usepackage{graphicx}
\usepackage{color}
\usepackage{amsmath}
\usepackage{amssymb}

\begin{document}


\title{Electron Heating During Magnetic Reconnection: A Simulation Scaling Study} 



\author{M. A. Shay}
\email[]{shay@udel.edu}
\author{C. C. Haggerty}

\affiliation{Bartol Research Institute, Department of Physics and
  Astronomy, University of Delaware, Newark, DE 19716, USA}

\author{T. D. Phan}
\affiliation{Space Sciences Laboratory, University of California, Berkeley, California 94720, USA}

\author{J. F. Drake}
\affiliation{Institute for Research in Electronics and Applied
  Physics, University of Maryland, College Park, Maryland 20742, USA}

\author{P. A. Cassak}
\affiliation{Department of Physics and Astronomy, West Virginia University, Morgantown, WV 26506, USA}

\author{P. Wu}
\affiliation{Bartol Research Institute, Department of Physics and
  Astronomy, University of Delaware, Newark, DE 19716, USA}
\affiliation{School of Mathematics and Physics, Queen's University, Belfast, United Kingdom}

\author{M. Oieroset}
\affiliation{Space Sciences Laboratory, University of California,
  Berkeley, California 94720, USA}

\author{M. Swisdak}
\affiliation{Institute for Research in Electronics and Applied
  Physics, University of Maryland, College Park, Maryland 20742, USA}

\author{K. Malakit}
\affiliation{Department of Physics, Mahidol University, Bangkok, 10400, Thailand}


\date{\today}

\begin{abstract}
Electron bulk heating during magnetic reconnection with symmetric inflow conditions is examined using kinetic particle-in-cell (PIC) simulations. Inflowing plasma parameters are varied over a wide range of conditions, and the increase of electron temperature is measured in the exhaust well downstream of the x-line. The degree of electron heating\( _{}\) is well correlated with the inflowing Alfv\'en speed \(c_{Ar}\) based on the reconnecting magnetic field through the relation \(\Delta T_e = 0.033 \,m_i\,c_{Ar}^2\), where \(\Delta T_{e}\) is the increase  in electron temperature.
 For the range of simulations performed, the heating shows almost no correlation with inflow total temperature \(T_{tot} = T_i + T_e\) or plasma \(\beta\). An out-of-plane (guide) magnetic field of similar magnitude to the reconnecting field does not affect the total heating, but it does quench perpendicular heating, with almost all heating being in the parallel direction. These results are qualitatively consistent with a recent statistical survey of electron heating in the dayside magnetopause (Phan et al, Geophys. Res. Lett., 40, doi:10.1002/grl.50917, 2013), which also found that \(\Delta T_e\) was proportional to the inflowing Alfv\'en speed. The net electron heating varies very little with distance downstream of the x-line. The simulations show at most a very weak dependence of electron heating on the ion to electron mass ratio. In the antiparallel reconnection case, the largely parallel heating is eventually isotropized downstream due a scattering mechanism such as stochastic particle motion or instabilities. The simulation size is large enough to be directly relevant to reconnection in the Earth's magnetosphere, and the present findings may prove to be universal in nature with applications to the solar wind, the solar corona, and other astrophysical plasmas. The study highlights key properties that must be satisfied by an electron heating mechanism: (1) Preferential heating in the parallel direction; (2) Heating proportional to \(m_i\,c_{Ar}^2\); (3) At most a weak dependence on electron mass; and (4) An exhaust electron temperature that varies little with distance from the x-line. \end{abstract}


\maketitle 

\section{Introduction }

Magnetic reconnection is a universal plasma process which converts stored magnetic energy into particle energy. The process is believed to be important in many astrophysical, solar, geophysical, and laboratory contexts. An important unresolved problem in reconnection research is to understand what controls electron energization in reconnection exhausts.  Past investigations have explored  suprathermal electron energization, both 
observationally [e.g., Ref.~\onlinecite{Baker76,Hoshino01b,Oieroset02,Chen08,Retino08}] and theoretically [e.g., Ref.~\onlinecite{Hoshino01b,Pritchett06,Drake06b,Egedal10}]. However, an even more basic problem is the reconnection associated thermal heating of electrons. By thermal heating, we mean heating of the core population and not the energetic tail of the distribution. Space observations suggest that the degree of thermal heating depends on plasma parameters. Strong heating is typically observed in reconnection exhausts in Earth's magnetotail\cite{Angelopoulos92}, while much weaker heating occurs in magnetopause\cite{Gosling06,Phan13} and solar wind exhausts\cite{Gosling07,Pulupa14}.

These disparate space observations may be consistent with the heating being primarily controlled by inflow conditions.
In a recent statistical observation study\cite{Phan13}, the degree of electron bulk heating in asymmetric reconnection exhausts at the Earth's magnetopause  was best correlated with the asymmetric outflow velocity\cite{Cassak07d,Swisdak07} \( C_{A-{\rm asymm}}^2\). A best fit to the data produced the empirical relation: \(\Delta T_e = M_{Te} \;m_i\, C_{A-{\rm asymm}}^2\), where \(M_{Te}\) is a constant with \(M_{Te} = 0.017,\) the ``\(\Delta\)''refers to the change in temperature from the magnetosheath inflowing plasma and \(T_e\) is related to the trace of the full electron temperature tensor \({\bf T}_e\) as  \(T_e = {\rm Tr}\, [{\bf T}_e] /3\).  The linear dependence of the heating indicates that the heating is proportional to the inflowing magnetic energy per proton-electron pair. It was also found in that study that perpendicular heating is substantially reduced in the presence of a strong guide field. 

Simulation case studies have examined electron temperatures and distributions during reconnection, finding that heating and associated anisotropies can be generated due to many mechanisms such as acceleration in the reconnection electric field, turbulent waves excited by Hall electric currents, betatron acceleration, Fermi reflection on curved moving field lines, and trapped electron populations due to parallel electric 
fields [e.g., Ref.~\onlinecite{Horiuchi97,Hoshino01a,Swisdak05,Fujimoto06b,Drake06b,Chen08b,Le09,Egedal12}].
A more recent kinetic PIC simulation study found that the dominant energization mechanism was Fermi reflection for nearly antiparallel reconnection and both Fermi reflection and parallel electric fields for stronger guide fields\cite{Dahlin14}.
A laboratory analysis of reconnection found that electrons are primarily energized close to the x-line with this energy transferred into the exhaust via heat conduction\cite{Yamada14}. In terms of theory and modeling, it is currently unclear how different reconnection conditions modify the magnitude of the electron heating and the heating mechanism. What is currently needed is a systematic simulation study of the degree of thermal electron heating in the exhaust region of magnetic reconnection and how it depends on a variety of inflow parameters. Such a study will directly test whether simulations can reproduce results consistent with observations, and will provide a testbed for determining the ultimate cause of the electron heating. 

We perform a series of fully kinetic particle-in-cell simulations examining the scaling of the electron heating for a range of inflow conditions and parameters. In this initial study, we choose first to focus on the simpler case of symmetric  reconnection, which will provide context when the more complicated asymmetric reconnection is examined at a later date. Even so, the key findings in terms of scaling with the inflow Alfv\'en speed \( (\,\Delta T_e \propto \;m_i\, C_{A{\rm in}}^2\,) \) and the anisotropy of heating  are remarkably similar to the asymmetric reconnection observations\cite{Phan13}, suggesting that this scaling is generic to reconnection.

The results have the following implications for an electron heating mechanism:
 (1) Preferential heating in the parallel direction; (2) Heating proportional to \(m_i\,c_{Ar}^2\), where   \(c_{Ar}\) is the inflow Alfv\'en speed based upon the reconnecting magnetic field; (3) At most a weak dependence on electron mass; and (4) An exhaust electron temperature that varies little
with distance from the x-line. 

The present paper is organized as follows. In Section~\ref{Sec:Theory}, the
theoretical context for electron heating during magnetic reconnection is
examined. Section~\ref{Sec:SimInfo} describes the numerical simulations in this study. Section~\ref{Sec:TypicalSimResults} gives  an example simulation. Section~\ref{Sec:DetermineHeating} describes how the
degree of electron heating is determined from the simulations. Section~\ref{Sec:ScalingHeating}
describes the scaling of the heating. Section~\ref{Sec:MassRatioDependence} examines the effect of electron
to ion mass ratio on the heating.  Section~\ref{Sec:DiscConclusions} is the discussion and conclusion
section. 

\section{\label{Sec:Theory}Theory}

In order to give context to the analysis of simulation data, we examine the heating using Sweet-Parker reconnection theory (a control volume analysis). For full generality, we first perform the analysis on asymmetric reconnection and then take the symmetric limit for application to this study. Our analysis is similar to previous Sweet-Parker analyses of asymmetric reconnection \cite{Cassak07d,Birn2010}.

\begin{figure}[p]
\begin{center}
\includegraphics[width=3.5in]
{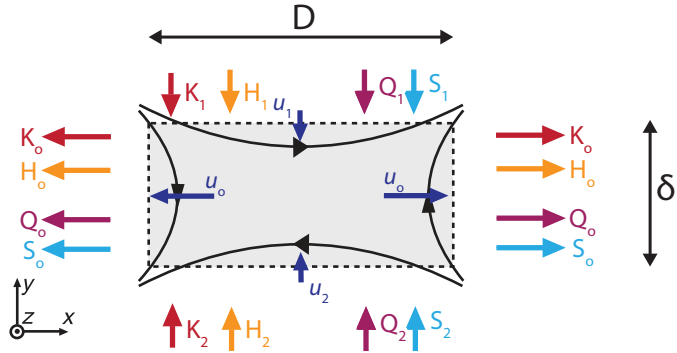}
\end{center}
\caption{Schematic of the energy fluxes into and out of the diffusion region for asymmetric reconnection. Subscripts ``1'' and ``2'' denote different inflowing quantities, and subscript ``o'' denotes outflowing quantities. \(u\) is bulk flow velocity, \(K\) is bulk flow energy flux, \(H\) is enthalpy flux, \(Q\) is heat flux, and \(S\) is electromagnetic Poynting flux. Adapted from [Eastwood et al., 2013]\cite{Eastwood13}. }
\label{Fig:Schematic}
\end{figure}

Figure~\ref{Fig:Schematic} shows a schematic of the energy fluxes into and out
of the diffusion region. \(u\) denotes bulk flow velocities. \(\delta\) is the
width of the outflow exhaust and \(D\) is the width of the inflow region. \(S\)
is Poynting flux, \(H\) is enthalpy flux, \(K\) is bulk fluid kinetic energy flux,
and \(Q\) is heat flux. The inflowing conditions on the two sides have subscript
``1'' and ``2,'' and the outflowing quantities have subscript ``o.'' Conservation
of energy requires: 

\begin{equation}
D\;\left( S_1 + S_2 + H_1 + H_2 + K_1 + K_2 + Q_1 + Q_2 \right)
 \approx 2\,\delta\; \left( S_{o} + H_{o} + K_{o} + Q_{o} \right).
\end{equation}


Ignoring the typically small incoming kinetic energy \(K_1\) and \(K_2\) and heat flux \(Q_1\) and \(Q_2\), this equation can be rewritten:
\begin{equation}
(S_1 + S_2) D\; \approx\; 2\delta S_{o} + [\;2\delta H_{o} - D (H_1 + H_2)\;]
\\+ 2\delta K_{o} + 2\delta Q_{o}.
\end{equation}

Dividing by the incoming Poynting flux yields \mbox{\( 1 \approx R_S + R_H + R_K + R_Q \),} where each \(R\) term represents the fractional amount of energy (relative to the converted magnetic energy) which leaves the diffusion region as each energy type. This study is focused the amount of energy going into heating, which is directly related to the enthalpy flux leaving the diffusion region. 
\begin{equation}
R_H = \frac{2\,\delta H_{o} - D(H_1 + H_2)}{D(S_1 + S_2)}.
\end{equation}
This fractional enthalpy flux can be broken up into contributions from the ions and electrons as \(R_{H}=R_{Hi}+R_{He}. \) For this study, we focus on the fractional electron enthalpy flux \(R_{He}\) which is written using the definition of enthalpy as:
\begin{equation}
R_{He} = \frac{\Gamma \left[\; 2 \delta u_o P_{eo} - D (u_1 P_{e1}  + u_2 P_{e2} ) \;\right]   }
{\frac{c}{4 \pi} E_z (B_1 + B_2) D},
\end{equation}
where \(\Gamma \equiv \gamma/(\gamma-1)\), with \(\gamma\)  the ratio of specific heats. It is assumed that the inflowing \(\Gamma\) is equal to the outflowing \(\Gamma\), the applicability of which will be discussed in Section~\ref{Sec:DiscConclusions}. Note that we have written \(S_1 = (c/4\pi)\,E_z \,B_1,\)  with a similar relation for \(S_2\). By doing so, we have discounted any Poynting flux associated with the out-of-plane (guide) magnetic field along \(\hat{z}\). Because little \(B_z\) energy is expected to be released in the diffusion region, this is a good approximation.

 Using continuity, \(2 \delta\, n_o u_o \approx D\, (n_1 u_1 + n_2 u_2)\), along with \(u_1 = cE_z/B_1\) and \(u_2 = cE_z/B_2\), yields a relation for \(R_{He}\):
\begin{equation}
R_{He} \approx \frac{\Gamma (T_{e0} - T_{ein} )}{m_i u_o^2},
\end{equation}
with the definitions:
\begin{equation}
T_{ein} = \frac{T_{e1} n_1 B_2 + T_{e2} n_2 B_1}{n_1 B_2 + n_2 B_1}
\end{equation}
\begin{equation}
u_o^2 = \frac{B_1 B_2}{4 \pi m_i} \; 
\frac{B_1 + B_2}{n_1 B_2 + n_2 B_1}
\end{equation}
The form of \(T_{ein}\) results from the fact that \(T_{e1}\) and \(T_{e2}\) are convected into the diffusion region with different velocities; it is the temperature of the outflowing plasma if there were only mixing and no heating. Therefore, to measure the actual change in thermal energy requires \(T_{eo}-T_{ein}\). Note that \(u_o\) is the outflow velocity for asymmetric reconnection\cite{Cassak07d,Swisdak07}.

\(m_{i}\,u_{o}^2\) represents the available inflowing magnetic free energy per proton-electron pair, which can be shown by dividing the incoming Poynting flux by the inflowing particle density flux: 
\begin{equation}
\frac{(S_1 + S_2) \,D}{(n_1 u_1 + n_2 u_2) \,D} = \frac{B_1 B_2}{4\pi}\;\frac{B_1 + B_2}{n_1 B_2 + n_2 B_1} = m_i u_0^2
\end{equation}

Note that the simulations in this study and observations of reconnection are not in thermodynamic equilibrium, with non-gaussian distribution functions and multiple beams.
For that reason there is uncertainty as to the most appropriate value of \(\Gamma\) to use for the outflowing plasma. We focus therefore simply on the ratio: \begin{equation}
M_{Te} = \frac{T_{eo} - T_{ein}}{m_i u_o^2}\label{Eqn:MTe-asymm}.
\end{equation}

  \(M_{Te}\) is a quantity that can be determined in a straightforward manner from each reconnection simulation, and is proportional to the amount of inflowing magnetic energy converted into electron heating.  An important question regards the variation of \(M_{Te}\) with changing inflowing parameters.   It seems quite plausible that the percentage of magnetic energy converted to electron heating during magnetic reconnection would have a dependence on inflow conditions. If, on the other hand, \(M_{Te}\) is a constant for a wide range of inflowing parameters, then the percentage of inflowing magnetic energy converted into electron heating is  a constant.

In the symmetric reconnection limit, Eq.~\ref{Eqn:MTe-asymm} simplifies to \mbox{\(M_{Te} = (T_{eo} - T_{ein})/(m_i c_{Ar}^2)\)}, where \(c_{Ar}\) is the Alfv\'en speed of the inflowing plasma based on the reconnecting magnetic field.

Another point to emphasize when studying the energy budget of reconnection regards the percentage of free energy converted to bulk outflows \(R_{K}\). The Poynting flux of energy represents a ``magnetic enthalpy'' [e.g., Ref.~\onlinecite{Priest00}]. and therefore contains twice the energy needed to accelerate the outflowing plasma to \(u_{o}\), i.e., dividing outflow kinetic energy flux for a velocity \(u_{o}\) by the incoming Poynting flux yields:
\begin{equation}
R_K = \frac{\left( \frac{1}{2}\,m_i \,n_o \,u_o^3\right)\; 2 \delta}{(S_1 + S_2)\,D} = \frac{1}{2}.
\end{equation}
Even if 50\% of the available inflowing magnetic energy is converted to bulk outflow energy, there will still be ample remaining magnetic energy to simultaneously heat the plasma. 
\section{\label{Sec:SimInfo}Simulation Information}

We use the parallel PIC code P3D\cite{Zeiler02}  to perform simulations in 2.5 dimensions of collisionless antiparallel reconnection. In the simulations, magnetic field strengths and particle number densities are normalized to arbitrary values \(B_{0}\) and \(n_{0}\), respectively.
 Lengths are normalized to the ion inertial length \(d_{i0}=c/\omega_{pi0}\) at the reference density \(n_{0}\).  Time is normalized to the ion cyclotron time \(\Omega_{ci0}^{-1}=(eB_{0}/m_{i}c)^{-1}.  \) Speeds are normalized to the Alfv\'en speed \(c_{A0}=\sqrt{B_{0}^{2}/(4\pi\,m_{i}\,n_{0})}\). Electric fields and temperatures are normalized to \(E_{0}=c_{A0}B_{0}/c\)  and \(T_{0}=m_{i}c_{A0}^{2}\), respectively. The coordinate system is a generic ``simulation coordinates,'' meaning that the reconnection outflows are along \(\hat{x}\) and the inflows are along \(\hat{y}\), as illustrated in Figure~\ref{Fig:Schematic}. 

Simulations are performed in a periodic domain with size and grid scale varied based on  simulation and inflow parameters; upstream densities of n = 1.0, 0.2 and 0.04 have \(L_{x} \times L_y = 204.8 \times 102.4, \;204.8 \times 102.4,\) and \(409.6 \times 204.8\) respectively. There are three mass ratios \(m_{i}/m_{e}=25, 100,400, \) with  grid scales \(\Delta_{x}=\Delta_{y}=0.05,\, 0.025, \,0.0125\)  and speed of light \(c = 15,  \,30,\,\ 40\) respectively. The initial conditions are a double current sheet\cite{Shay07}.
A small magnetic perturbation is used to initiate reconnection.
Each simulation is evolved until reconnection reaches a steady state, and then during the steady-state period the simulation data is time averaged over 100 particle time steps, which is typically on the order of 50 electron plasma wave periods \(\omega_{pe}^{-1}\). 

In order to examine the effect of inflowing plasma conditions on electron heating, the initial simulation inflow parameters are varied over a range of values shown in Table~\ref{Tbl:Sims}. Variations in parameters are: reconnecting magnetic field Br between \(1/\sqrt{5}\) and \(\sqrt{5}\), density \(n_{in}\) between 0.04 and 1.0, inflowing electron temperature \(T_{e}\) between 0.03 and 1.25, and \(T_{i}/T_{e}\) between 1 and 9.
Simulations have either no guide field (anti-parallel reconnection) or a guide field \(B_{g}=B_{r}\) (magnetic shear angle of \(90^{\circ}\)).
The initial upstream reconnection Alfv\'en speed has values \(c_{Ar}^2 \equiv B_r^2/n_{in} = 1.0,\; 5.0,\; 17.0,\; {\rm and}\; 25.0\). The plasma total \(\beta\) ranges from 0.06 to 5.0. 

Note that for the purpose of  connection with the Phan et al., 2013 magnetosheath inflow conditions\cite{Phan13}, many of the \(\beta\) values are substantially larger than typically used in generic kinetic PIC simulation studies. For example, the GEM challenge study\cite{Birn01} had inflow \(\beta=0.2.\)

\section{\label{Sec:TypicalSimResults}Simulation Example }
An overview of the reconnecting system is shown for run 46 in Figure~\ref{Fig:overview}: (a) \(V_{ez}\) and (b)
\(V_{ex}\) with magnetic field lines, (c) \(B_{z}\), (d) \(T_{e||}\), (e)
\(T_{e\perp}\), and (f) \(T_{e}=(T_{e||}+2\,T_{e\perp})/3\). Note that plots
(d), (e), and (f) are on the same color scale to facilitate comparison.
The out-of-plane electron flow is typical for anti-parallel reconnection, with flows near the x-line comparable to the electron Alfv\'en speed, and weaker flows near the separatrices and downstream of the x-line.
The electron outflow shows the super-Alfv\'enic electron jets
associated with the outer electron diffusion
region\cite{Shay07,Karimabadi07}, as well as the parallel electron
flows near the separatrices associated with Hall currents. The
out-of-plane magnetic field has the typical quadrupolar structure.

The heating of the electrons is evident in Figure~\ref{Fig:overview}d-f. There is strong electron parallel heating in the exhaust of the reconnection region. The perpendicular heating is  localized very close to the midplane near the x-line but broadens to include the whole exhaust region downstream.
In terms of the electron heating, we define the ``near exhaust'' \(( 45 \lesssim x \lesssim 75 )\) as the region with little perpendicular heating away from the midplane, and the ``far exhaust''    \(\left[ \,( 25 \lesssim x \lesssim 45 )\;\; {\rm and} \;\; ( 75 \lesssim x \lesssim 90 )\, \right  ]\) as the regions downstream of that but before the edge of the reconnection jet front (in the past called the  ``dipolarization front''). The near exhaust is therefore associated with strong electron temperature anisotropy, while the temperature is more isotropic in the far exhaust. 

A striking property of the heating in Figure~\ref{Fig:overview}f is that both the near and far exhausts are characterized by a nearly constant \(T_{e}\). The constancy of \(T_{e}\) with distance downstream of the x-line implies that electrons are continually being heated in the exhaust, with heating being just enough to bring the inflowing unheated plasma up to the exhaust temperature.
The lack of perpendicular heating in the near exhaust implies that the heating mechanism first heats electrons along the parallel direction, with this parallel energy later being scattered into the perpendicular direction.

\section{\label{Sec:DetermineHeating}Determination of Heating}

We determine the downstream heating by examining a slice along \(y\) in the exhaust at the following downstream distances: (1) \(n_{in}\) = 0.2, distance = \(20\,d_{i0}\), (2) \(n_{in}\) = 1.0, distance = \(9\,d_{i0}\), and (3) \(n_{in}\) = 0.04, distance = \(45\,d_{i0}. \) Normalized to the ion inertial length in the inflow region, these distances are all the same.
All data used in the analysis of electron heating has been time averaged over 100 time steps, which is typically about 50 electron plasma wave periods \(\omega_{pe}^{-1}\).

Figure~\ref{Fig:DetermineHeating} shows slices of data along \(y\) for the simulation described in Figure~\ref{Fig:overview}:
(a) Magnetic fields, (b) Ion flows, (c) Electron Flows, (d) Electron Temperature, which shows typical exhaust properties for this type of reconnection.
In (a) the quadrupolar Hall magnetic fields are evident, filling most of the exhaust region.  In (b), the ion exhaust region is evident in red. Electron flows in the \(x\) direction in red (c) show the super-Alfv\'enic electron outflows as well as the parallel flows towards the x-line near the separatrices. Plots of \(T_{e},\,T_{e||},\) and \(T_{e\perp}\)   are shown in (d).   There is a sharp drop of \(T_{e||}\) and a sharp rise of \(T_{e\perp}\) near the midplane, while \(T_{e}\) stays relatively constant. Evidently, the electron thermal energy is simply being transferred between the perpendicular and parallel directions.

To determine the heating occurring in the outflow exhaust, we calculate the spatial average of the temperature in the exhaust \(\overline{T_e}\), and subtract the average inflow temperature \(T_{ein}\), yielding  \(\Delta T_e = \overline{T_e} - T_{eup}\).
We calculate the anisotropic heating \(\Delta T_{e||},\,\Delta T_{e\perp}\), and the total electron heating \(\Delta T_e = (\Delta T_{e||} + 2\,\Delta T_{e\perp})/3\).
For Figure~\ref{Fig:DetermineHeating}, the two upstream regions which determine the inflow values are shown with the vertical dotted lines. The exhaust region boundaries in this case are shown by the vertical dashed lines. In addition, the standard deviation of the temperature in the exhaust region is determined.

\section{\label{Sec:ScalingHeating}Scaling of Heating}

\begin{figure}[p]
\begin{center}
\includegraphics[width=3.4in]{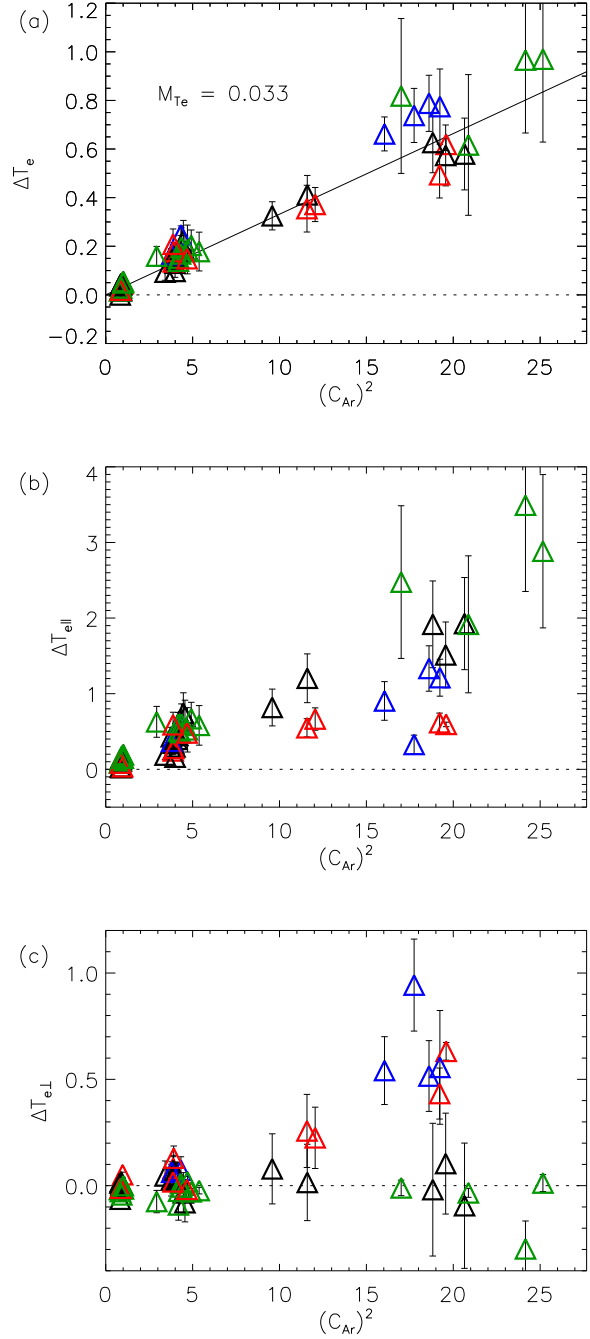}

\end{center}
\caption{(a) \(\Delta\,T_{e}\), (b) \(\Delta\,T_{e||}\),  and (c) \(\Delta\,T_{e\perp}\)
versus \(c_{Ar}^{2}\) . Standard deviations of the averaging shown as error
bars. Color of symbol represents type of run: (green) \(m_{i}/m_{e}=25\)
with guide field; (blue) \(m_{i}/m_{e}=25\), antiparallel, \(\beta<0.6;\)
(black) \(m_{i}/m_{e}=25\), antiparallel, \(\beta\geq0.6;\)
  (red) \(m_{i}/m_{e}=100.\) }
\label{Fig:Scaling}
\end{figure}

The scaling of the heating for 56 simulations is shown in Figure~\ref{Fig:Scaling}: (a) \(\Delta\,T_{e}\), (b) \(\Delta\,T_{e||}\),  and (c) \(\Delta\,T_{e\perp}\)
versus \(c_{Ar }^{2},\) where \(c_{Ar}\)  is the Alfv\'en speed (using the reconnecting magnetic field) based upon the average upstream conditions determined from each run (as shown in Figure~\ref{Fig:DetermineHeating}d). The colors of the symbols represent some important properties of each run: (green) \(m_{i}/m_{e}=25\)
with guide field; (blue) \(m_{i}/m_{e}=25\), antiparallel, \(\beta<0.6;\)
(black) \(m_{i}/m_{e}=25\), antiparallel, \(\beta\geq0.6;\)
  (red) \(m_{i}/m_{e}=100.\) The standard deviations of the temperature are shown as error bars for each data point.

As discussed in Section~\ref{Sec:Theory}, for each simulation the
percentage of magnetic energy converted to electron heating is
proportional to: \(M_{Te} = \Delta T_e/(m_i\,c_{Ar}^2)\).  In
Figure~\ref{Fig:Scaling}a, \(\Delta T_{e}\) for each simulation is
plotted versus \(m_{i}c_{Ar}^{2}\).  The data roughly follows a straight line, meaning
that the percentage of magnetic energy converted into electron heating
is approximately constant across the simulations. The best fit line through
the origin, fitting \(\Delta T_{e}=M_{Te}\,c_{Ar}^{2}\), yields
\(M_{Te}=0.033, \) which is about twice the slope from Phan et al.,
2013.  What is striking is the universality of the scaling of electron
temperature, independent of guide field and \(\beta\), which vary
considerably over the 56 runs.

To verify that parameters such as \(\beta\) and temperature are not playing a primary role in determining the heating, in Figure~\ref{Fig:noBeta-noT}  we plot the dependence of electron heating on  the inflowing values of (a) \(\beta_{r}\) and (b) \(T_{tot}=T_{i}+T_{e}. \) \(\beta_{r}\) is determined using the reconnecting magnetic field component.   Care must be taken in analyzing the results because the simulation space does not fill in all of parameter space.
We therefore organize the data points by the asymptotic upstream Alfv\'en speed:
(black) \(c_{Ar}^{2}=25; \)  (blue) \(c_{Ar}^{2}=14;\) (green) \(c_{Ar}^{2}=5;
\) and (magenta) \(c_{Ar}^{2}=1.\)  It may appear that there is some heating dependence on \(\beta_{r}\), with less heating for higher \(\beta_{r}\).
However, the color coding makes it clear that this dependence is likely due to the dearth of high \(\beta_{r}\) with high Alfv\'en speed simulations, which are computationally challenging to perform.
 It is clear that any affect on heating from \(\beta_{r}\) and
 \(T_{tot}\) plays at most a secondary role to the upstream Alfv\'en
 speed.

A different story emerges from the scaling of \(\Delta T_{e||}\) and \(\Delta T_{e\perp}\)  because the spatial structure of the anisotropy depends on \(\beta\).
  Examining heating in the exhaust at a fixed distance from the x-line leads to different measured anisotropies. Figure~\ref{Fig:Scaling}b and \ref{Fig:Scaling}c show the parallel and perpendicular heating, respectively.
Focussing first on the guide field cases written as green points, it is striking that there is no perpendicular heating in these cases. A surprise, however, is that several of the anti-parallel simulations exhibit this anisotropy also, with little or no perpendicular heating. The reason to separate the \(m_{i}/m_{e}=25\) cases into high \(\beta\) and low \(\beta\) becomes clear in Figures~\ref{Fig:Scaling}b and \ref{Fig:Scaling}c. For the high \(c_{Ar}\) cases, the guide field (green symbols) and the black symbols (higher \(\beta\)) show no perpendicular heating and greater parallel heating. This points to a faster isotropization closer to the x-line for the lower \(\beta\) simulations with \(m_{i}/m_{e}=25 \) as  well as all of the \(m_{i}/m_{e}=100\) cases.

\begin{figure}[p]
\begin{center}
\includegraphics[width=6.5in]{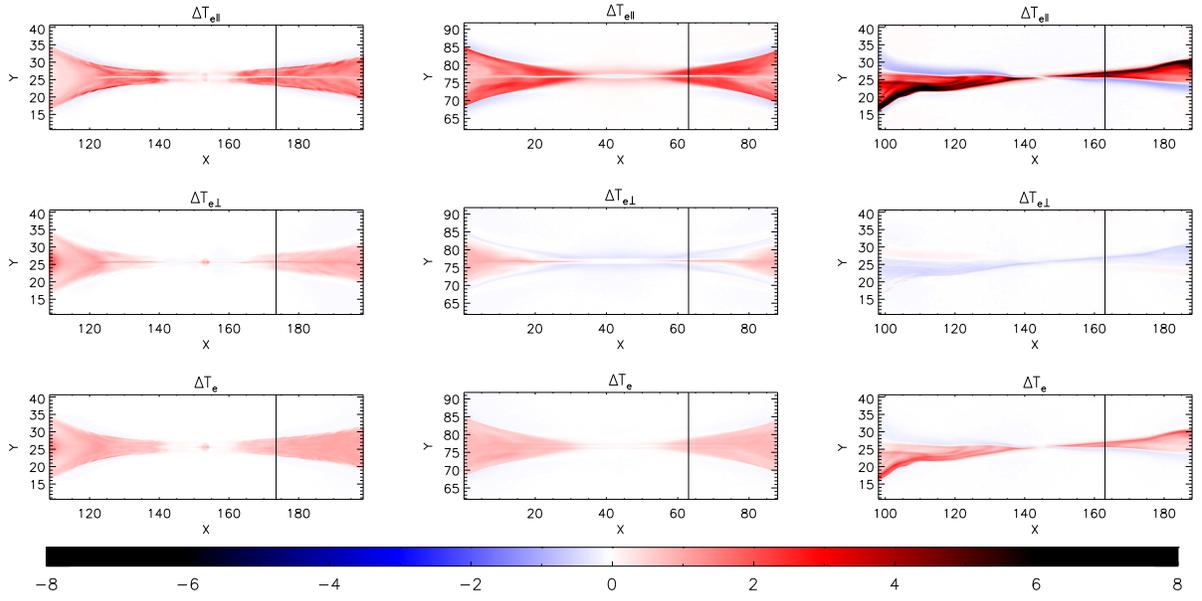}
\end{center}
\caption{Change in temperature relative to upstream value for three different runs highlighting the change in the character of the heating for the change in \(\beta\) and the change in guide field. All runs have \(m_{i}/m_{e}=25.\) (left) run 25  with no guide field and \(\beta=0.12;\) (middle)  run 33  with no guide field and \(\beta=0.6;\) (right) run 45 with guide field equal reconnecting field and \(\beta=0.3.\)}
\label{Fig:beta-dependence}
\end{figure}

Figure~\ref{Fig:beta-dependence} shows this difference in isotropization in more detail, where the
the change in electron temperature from the upstream values are shown for
\(m_{i}/m_{e}=25 \) cases with varying \(\beta\) and guide field: (left) run 25  with no guide field and \(\beta=0.12;\) (middle)  run 33 
with no guide field and \(\beta=0.6;\) (right) run 45 with guide field equal
reconnecting field and \(\beta=0.3. \) The vertical line in the figure shows for each run where the vertical slice was taken to determine the heating.

Focussing on the anti-parallel cases first (left and middle columns), both
show exhaust-filling total electron heating \(\Delta T_{e}\) which onsets
about \(10\,d_{i0}\) downstream of the x-line. As with run 46 in Figure~\ref{Fig:overview},
this average \(\Delta T_e\)  is relatively uniform beyond \(10\,d_{i0}.\)
Note that the leftmost simulation has just started to develop a secondary island. For both \(\beta\) values the onset of parallel heating occurs closer to the x-line than the perpendicular heating. However, for the lower \(\beta\) case, \(\Delta T_{e\perp}\)  becomes exhaust filling perhaps \(20\,d_{i0} \) downstream, whereas for the higher \(\beta\) case this does not occur until around \(30\,d_{i0}\) downstream. The lower \(\beta\) case is isotropizing faster than the higher \(\beta\) case.  

The reason for this behavior is that lower \(\beta\) cases exhibit stronger electron beaming relative to the electron thermal velocity and thus are much more susceptible to two-stream instabilities and electron hole formation\cite{Cattell05}. In Figure~\ref{Fig:beta-dependence}, these instabilities are apparent in \(\Delta T_{e||}\) for the low \(\beta\) case {} as spatial fluctuations   which onset simultaneously with the heating about \(10\,d_{i0}\) downstream
of the x-line. In contrast, the higher \(\beta\) case has a much smoother \(\Delta T_{e||}\), until around \(x=75\,d_{i0},\)  where oscillations become apparent. These may be due to a firehose-type instability, which isotropizes the electron temperature. 

The guide field case is fundamentally different from the anti-parallel cases. The heating in the exhaust is strongly asymmetric along the normal direction (along~\(y\)), and there is almost no \(\Delta T_{e\perp}.\) These findings provide  evidence that the heating mechanism or mechanisms first heat the electrons along the parallel direction which then scatters into the perpendicular direction.

\section{\label{Sec:MassRatioDependence}Mass Ratio Dependence of Heating}

An important question regards whether there is a mass dependence on the electron heating, as a realistic mass ratio is beyond the current supercomputer capabilities for a large scale statistical study such as this.
Clearly, from Figure~\ref{Fig:Scaling}a, any mass ratio dependence is  weak.  The \(m_{i}/m_{e}=100\)  cases do have slightly lower heating for the highest \(c_{Ar}\) values, but the difference is small.

\begin{figure}[p]
\begin{center}
\includegraphics[width=3.4in]{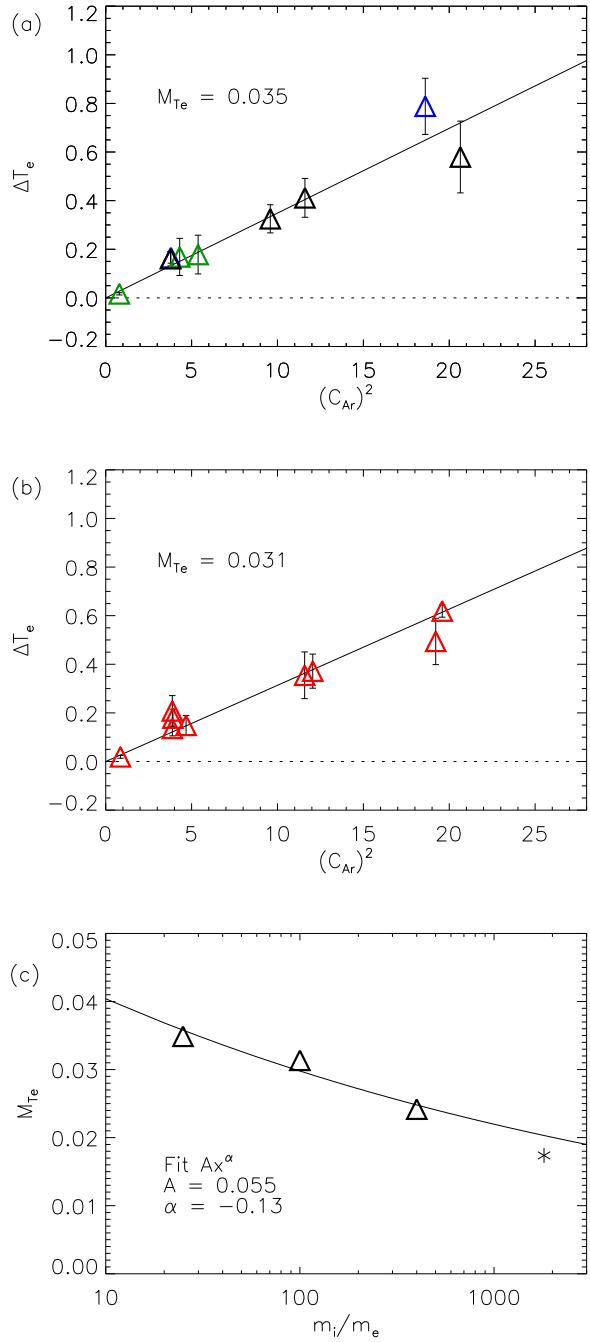}
\end{center}
\caption{Effect of Mass ratio on Electron Heating. (a) \(m_{i}/m_{e}=25\) and (b) \(m_{i}/m_{e}=100\) simulations with the same parameters except for mass ratio. (c) \(M_{Te}\) versus mass ratio. Note that the \(m_{i}/m_{e}=400\) point is from a single simulation.
The coloring of points in panels (a) and (b) uses the same convention as in Figure~\ref{Fig:Scaling}. The simulations used for this figure are shown in Table~\ref{Tbl:Sims} with a check mark in the ``\(m_{i}/m_{e}\) compare'' column.}\label{Fig:massratio}
\end{figure}

To put this difference on a more numerical basis, we compare \(M_{Te}\) for two different mass ratios. To make the comparison as straightforward as possible, we only compare simulations that have the same initial density, temperatures, and magnetic fields; these runs have a check mark in the ``\(m_{i}/m_{e} \) compare'' column in Table~\ref{Tbl:Sims}. Figure~\ref{Fig:massratio} shows \(\Delta T_{e}\) versus \(m_{i}\,c_{Ar}^{2}\) for (a) \(m_{i}/m_{e}=25\) and (b) \(m_{i}/m_{e}=100.   \) The coloring of data points uses the same convention as in Figure~\ref{Fig:Scaling}. There is a \(\approx 10\%\) difference in \(M_{Te} \) for the two mass ratios.

To provide a tentative scaling of heating versus mass ratio, we plot \(M_{Te}\) versus \(m_{i}/m_{e}\) in Figure~\ref{Fig:massratio}c and calculate the best fit curve with the   functional form \(A\:(m_{i}/m_{e\\ })^{\alpha}\).
Note that the \(m_{i}/m_{e}=400\) case  is a single simulation, run 56. A power law dependence with \(A=0.055\) and \(\alpha=-0.13\) is found, which as expected is a very weak dependence on mass ratio. 

Extending this fit to a realistic mass ratio of \(m_{i}/m_{e}=1836, \) we find \(M_{Te}=0.020.\) This value is much closer to the experimental value from Phan et al., 2013 of 0.017, which is plotted as an asterisk in Figure~\ref{Fig:massratio}. Thus, this weak mass ratio dependence is one possible explanation for the difference between the magnetopause observations findings and this simulation study. 

\section{\label{Sec:DiscConclusions}Discussion and Conclusions}

A systematic kinetic-PIC simulation study of the effect of inflow parameters on the electron heating due to magnetic reconnection has been performed.We find that electron heating is well characterized by the inflowing Alfv\'en speed through the relation \(\Delta T_e = M_{Te}\;m_i c_{Ar}^2\), where \(M_{Te}\) is a constant of 0.033.  For the range of inflow parameters  performed, the heating shows almost no correlation with total temperature \(T_{tot}=T_{i}+T_{e}\) or plasma \(\beta\). A guide field of similar magnitude to the reconnecting field quenches perpendicular heating, with almost all heating being in the parallel direction.
These findings are qualitatively consistent with a recent observational study of electron heating\cite{Phan13}, which also found that \(\Delta T_e\) was proportional to the inflowing
Alfv\'en speed. A significant point regarding the simulation/observation
comparison is that the observational study examined asymmetric
inflow conditions while the simulations were of symmetric
reconnection. Such an agreement implies that there may be a
generic heating mechanism at work, and makes a case for the
universality of the results of this study and the observational
study.   

An important question regarding magnetic reconnection is the ultimate fate of the released magnetic energy, i.e., the determination of the \(R\) values described in Section~\ref{Sec:Theory}. MHD theory predicts that significant amounts of the released magnetic energy is converted to thermal energy, even in the incompressible limit\cite{Birn2010}. The percentage of inflowing Poynting flux converted into electron enthalpy  flux is given as \(R_{He} = \Gamma \;\Delta T_e \,/\, (m_i c_{Ar}^2)\), as reviewed in Section~\ref{Sec:Theory}. For an isotropic plasma, the average \(M_{Te}=0.033\) in this study corresponds to the following percentage of inflowing Poynting flux converted to electron enthalpy flux: \(R_{He} = 5/2 \,(0.033) = 0.083\) or 8.3\%. The Phan et al., 2013 observations give  \(R_{He} = 5/2 \,(0.017) \approx 0.043\), or 4.3\%. 

There is uncertainty in these percentages because both observations and kinetic PIC simulations exhibit temperature anisotropy in the exhaust (in the simulations the inflowing plasma is nearly isotropic).  In a kinetic plasma with a pressure tensor \({\bf P}\), the general form for the ``kinetic'' enthalpy flux is \({\bf H}_k = (3/2)\,{\bf u}\,P + {\bf u} \cdot {\bf P}\),  where \(P \equiv {\rm Tr}[{\bf P}]\,/\,3\).  If \(T_{e||}\gg T_{e\perp},\) for example,  the enthalpy flux along the magnetic field line would be 9/5 larger than the isotropic enthalpy flux, while the flux perpendicular to the field line would be 3/5 of the isotropic case.
However, a preliminary analysis was performed examining both antiparallel and guide field cases in this study, and it was found that the integrated kinetic enthalpy flux across the exhaust was nearly equal to the predicted isotropic enthalpy flux.

The primary quantitative difference between this study and the observations is the value of \(M_{Te}\), which for the simulations is approximately twice the value of the observations. The simulations do show a weak dependence on the electron mass with \(\Delta T_{e} \approx 0.055\,(m_i/m_e)^{-0.13}\), which when extrapolated to a realistic mass ratio gives \(M_{Te}\approx0.020,\) which is quite close to the \(M_{Te}=0.017\) seen in the magnetopause observations\cite{Phan13}. This would suggest that \(M_{Te}\) is truly a universal feature, as the reconnection observations were for asymmetric reconnection, while these simulations are symmetric. While this finding is interesting, there are significant uncertainties as to the mass ratio scaling, as well as many other possible explanations for the quantitative difference between simulations and observations: 2D versus 3D, symmetric versus asymmetric, and observational uncertainties such as distance from the x-line, to name a few. 

The relatively small electron enthalpy percentages for the simulations and
observations are consistent with the outoing flux of
energy being dominated by ion enthalpy flux, as seen in hybrid simulations\cite{Aunai11}
and satellite observations in the Earth's magnetotail\cite{Eastwood13}. A
recent laboratory study\cite{Yamada14} of reconnection found that a magnetic
energy inflow rate of \(1.9\pm0.2\, \text{MW}\) resulted in a change of electron
thermal energy of \(0.26 \pm 0.1\,\text{MW}\), which represents a conversion
rate of around 14\%. However, comparison of this percentage with our simulation
results  is
complicated because some aspects of the analysis methods for the laboratory study and our simulation study are different. For example,  unlike our quasi-steady analysis, the  laboratory experiment showed significant
time dependence which was included in the energy conversion rate. 

In all simulations, the heating in the exhaust region near the x-line is initially only in the parallel direction. For some cases, this parallel heating ultimately isotropizes at distances farther from the x-line. This finding implies that the heating mechanism primarily heats the plasma parallel to the magnetic field.

The isotropization of the parallel electron heating during antiparallel reconnection shows significant dependence on the upstream temperature and  \(\beta. \)  At lower \(\beta,\) streaming instabilities are stronger and thus the isotropization occurs closer to the x-line than for the higher \(\beta \) cases. 

A striking clue to the nature of the electron heating is that in the outflow exhaust \(T_{e}\) shows little variation with distance from the x-line. Because cold inflowing electrons are continually ejected into the exhaust, this implies that electrons are being continually heated even far from the x-line.

Although the mechanism for electron heating is uncertain at this point, the findings in this study constrain the possible mechanisms: (1) Heating proportional to \(m_{i}c_{Ar}^{2}\); (2) An exhaust electron temperature that varies little
with distance from the x-line; (3) A preferential heating in the parallel direction, and (4) At most a very weak dependence on electron mass on the order of \((m_{i}/m_{e})^{-0.13}.\) The parallel heating rules out betatron acceleration because it would preferentially heat the plasma along the perpendicular direction [e.g., Ref.~\onlinecite{Birn00}]. There exists a parallel potential in the exhaust region\cite{Egedal13}, which could lead to parallel heating through the generation of counterstreaming beams. On the other hand, Fermi-bounce heating through contracting magnetic field lines\cite{Drake06b,Dahlin14} also produces preferential parallel heating. A recent kinetic-PIC study\cite{Dahlin14}
which found  that electron energization was dominated by the Fermi reflection
term\cite{Drake06b} for nearly anti-parallel reconnection, and by parallel
electric fields and the Fermi mechanism in guide field reconnection. The physical mechanism of the electron heating mechanism will be a topic of a future study.

\begin{figure}[p]
\begin{center}
\includegraphics[width=3.4in]{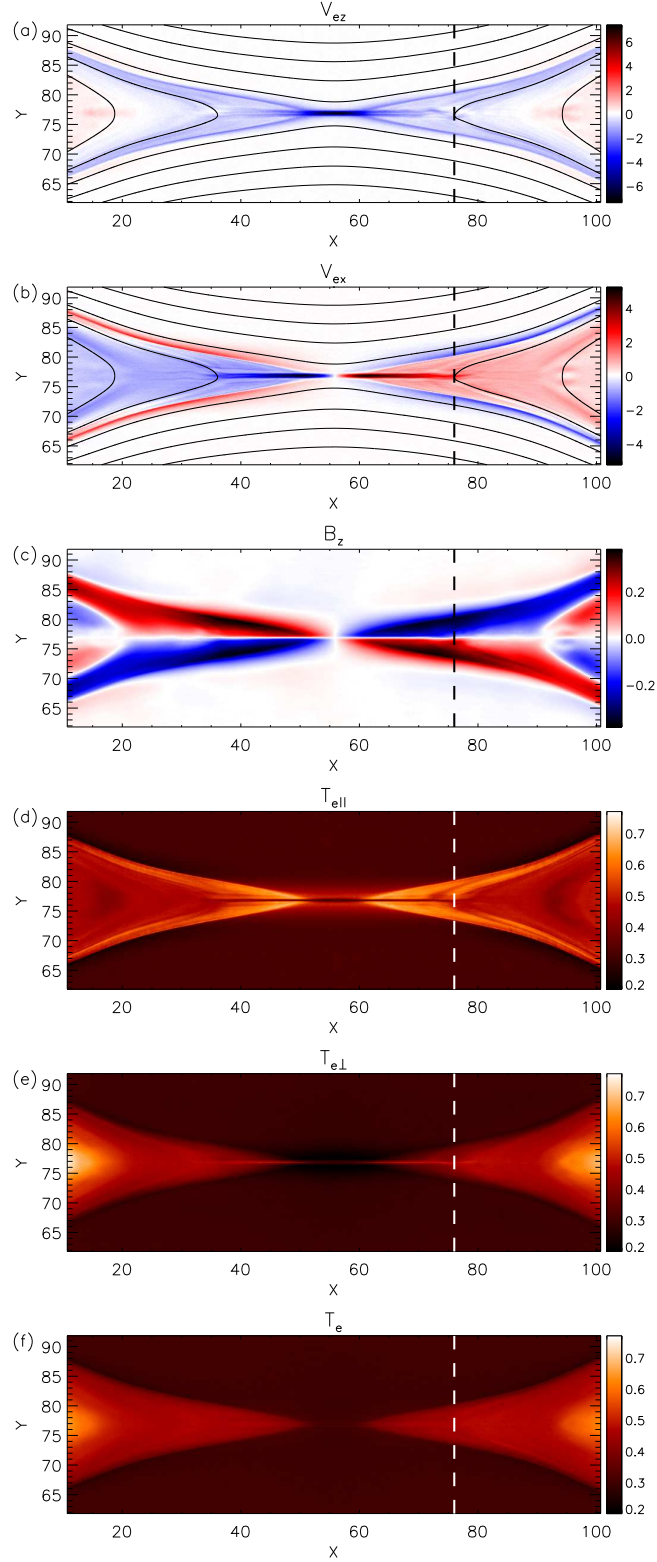}
\end{center}
\caption{Basic reconnection parameters for run 46. (a) \(V_{ez}\) and (b) \(V_{ex}\) with magnetic field lines, (c) \(B_{z}\), (d) \(T_{e||}\), (e) \(T_{e\perp}\), and (f) \(T_{e}=(T_{e||}+2\,T_{e\perp})/3\). Note that plots (d), (e), and (f) are on the same color scale for easy comparison. The vertical dashed lines show the location of the cut for Figure~\ref{Fig:DetermineHeating}.}
\label{Fig:overview}
\end{figure}

\begin{figure}[p]
\begin{center}
\includegraphics[width=3.4in]{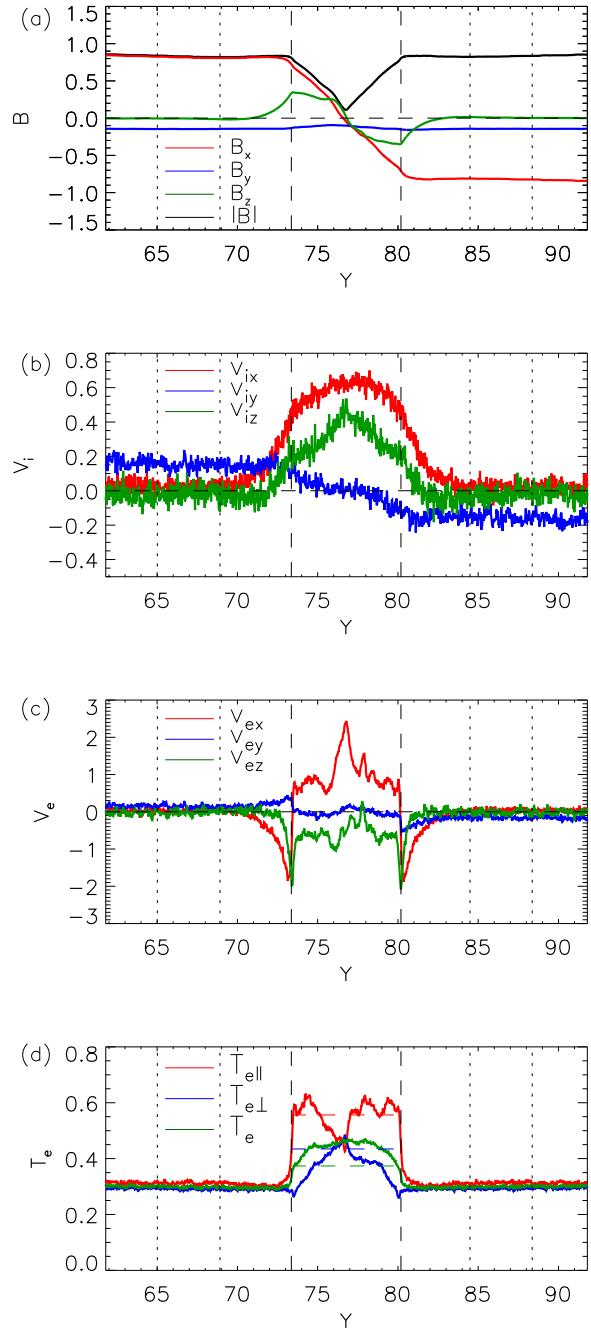}

\end{center}
\caption{Determination of electron heating. Slices taken at \(x = 76.0125\) in Figure~\ref{Fig:overview}. (a) Magnetic fields, (b) Ion flow velocities, (c) Electron flow velocities, (d) Electron temperatures. Dashed vertical lines show exhaust region and dotted vertical lines show inflow regions.}
\label{Fig:DetermineHeating}
\end{figure}

\begin{figure}[p]
\begin{center}
\includegraphics[width=3.4in]{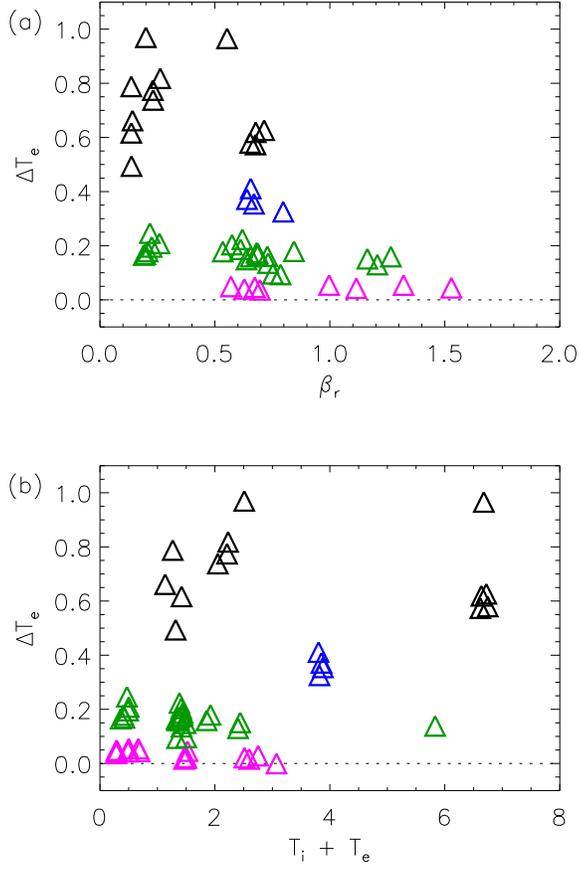}
\end{center}
\caption{Electron heating versus (a) \(\beta_{r}\) and (b) \(T_{tot}=T_{i}+T_{e}.\) \(\beta_{r} \) and \(T_{tot}\) are determined using the average values upstream when the electron heating is determined, as is described in Figure~\ref{Fig:DetermineHeating}. The color of the symbol refers to the asymptotic Alfv\'en speed in the upstream region using the asymptotic reconnecting field and density shown in Table~\ref{Tbl:Sims}: (black) \(c_{Ar}^{2}=25; \)  (blue) \(c_{Ar}^{2}=14;\) (green) \(c_{Ar}^{2}=5; \) and (magenta) \(c_{Ar}^{2}=1.\) }
\label{Fig:noBeta-noT}
\end{figure}

Energization and heating occurs naturally both at the x-line (e.g., Ref.~\onlinecite{Pritchett06} and references therein) and in the flux pile-up region at the edge of the exhaust\cite{Hoshino01b}. Electrons that travel close enough to the x-line to demagnetize can be accelerated along the reconnection electric field, causing heating and energization. In Figure~\ref{Fig:overview}, the width of this electron demagnetization region is a few \(d_{i0}\) along \(x\). With a reconnection rate \(E_{z}\approx0.12\) and with the change in flux from the x-line to the edge of the electron demagnetization region being about 0.04, it takes a magnetic field line a time of about 0.4 to reconnect and travel to the edge of this region.
Electrons that can propagate along a field line and enter this region during this time will be free accelerated to high velocities. With an upstream thermal velocity of around 7.0, only electrons within around \(3\,d_{i0}\) from this region will be free accelerated.
Therefore, the large majority of electrons in the simulation do not sample this inner region. If heating were only occurring very near the x-line, the electron temperature would be expected to decrease with distance from the x-line. 

 Regarding electron energization in the flux pileup region at the edge of the ion outflow exhaust, that region is transient in nature and is pushed downstream as the simulation progresses.
 In Figure~\ref{Fig:overview}, that region is around \(30\,d_{i0}\) downstream of the x-line. This heating study does not examine electrons that have passed through the flux pileup region. 

The applicability of this study for reconnection in physical systems is an important question, i.e., are the mechanisms of electron heating in the simulations likely to be similar to those found in actual physical systems? First, the consistency of these simulation results to the Phan et al., 2013 study is evidence for the relevance of the simulations. The findings of this study have been tested over a range of inflow conditions and ion to electron mass ratios. System size also plays an important role in the simulation relevance. While the simulations in this study are of sizes large enough to be applicable to reconnection in the magnetosphere, they are extremely small relative to distances in the solar wind and on the sun. However, the constancy of \(T_{e}\) with distance from the x-line in the simulations gives some credence to the idea that the simulation heating mechanism has converged with system size.


%
%

%

\begin{acknowledgments}
This research was support by the NASA Space Grant program at the University of Delaware; NSF Grants Nos. AGS-1219382 (M.A.S), AGS-1202330 (J. F. D),  and AGS-0953463 (P.A.C.); NASA Grants Nos. NNX08A083G--MMS IDS (T.D.P and M.A.S), NNX11AD69G (M.A.S.),   NNX13AD72G (M. A. S.), and  NNX10AN08A (P. A. C.). Simulations and analysis were performed at the National Center for Atmospheric Research Computational and Information System Laboratory (NCAR-CISL) and at the National Energy Research Scientific Computing Center (NERSC).  We wish to acknowledge support from the International Space Science Institute in Bern, Switzerland. 
\end{acknowledgments}

\newpage
\bibliographystyle{aipauth4-1}
\bibliography{bib}

\begin{thebibliography}{35}%
\makeatletter
\providecommand \@ifxundefined [1]{%
 \@ifx{#1\undefined}
}%
\providecommand \@ifnum [1]{%
 \ifnum #1\expandafter \@firstoftwo
 \else \expandafter \@secondoftwo
 \fi
}%
\providecommand \@ifx [1]{%
 \ifx #1\expandafter \@firstoftwo
 \else \expandafter \@secondoftwo
 \fi
}%
\providecommand \natexlab [1]{#1}%
\providecommand \enquote  [1]{``#1''}%
\providecommand \bibnamefont  [1]{#1}%
\providecommand \bibfnamefont [1]{#1}%
\providecommand \citenamefont [1]{#1}%
\providecommand \href@noop [0]{\@secondoftwo}%
\providecommand \href [0]{\begingroup \@sanitize@url \@href}%
\providecommand \@href[1]{\@@startlink{#1}\@@href}%
\providecommand \@@href[1]{\endgroup#1\@@endlink}%
\providecommand \@sanitize@url [0]{\catcode `\\12\catcode `\$12\catcode
  `\&12\catcode `\#12\catcode `\^12\catcode `\_12\catcode `\%12\relax}%
\providecommand \@@startlink[1]{}%
\providecommand \@@endlink[0]{}%
\providecommand \url  [0]{\begingroup\@sanitize@url \@url }%
\providecommand \@url [1]{\endgroup\@href {#1}{\urlprefix }}%
\providecommand \urlprefix  [0]{URL }%
\providecommand \Eprint [0]{\href }%
\providecommand \doibase [0]{http://dx.doi.org/}%
\providecommand \selectlanguage [0]{\@gobble}%
\providecommand \bibinfo  [0]{\@secondoftwo}%
\providecommand \bibfield  [0]{\@secondoftwo}%
\providecommand \translation [1]{[#1]}%
\providecommand \BibitemOpen [0]{}%
\providecommand \bibitemStop [0]{}%
\providecommand \bibitemNoStop [0]{.\EOS\space}%
\providecommand \EOS [0]{\spacefactor3000\relax}%
\providecommand \BibitemShut  [1]{\csname bibitem#1\endcsname}%
\let\auto@bib@innerbib\@empty
\bibitem [{\citenamefont {Angelopoulos}\ \emph {et~al.}(1992)\citenamefont
  {Angelopoulos}, \citenamefont {Baumjohann}, \citenamefont {Kennel},
  \citenamefont {Coroniti}, \citenamefont {Kivelson}, \citenamefont {Pellat},
  \citenamefont {Walker}, \citenamefont {Luhr},\ and\ \citenamefont
  {Paschmann}}]{Angelopoulos92}%
  \BibitemOpen
  \bibfield  {author} {\bibinfo {author} {\bibnamefont {Angelopoulos},
  \bibfnamefont {V.}}, \bibinfo {author} {\bibnamefont {Baumjohann},
  \bibfnamefont {W.}}, \bibinfo {author} {\bibnamefont {Kennel}, \bibfnamefont
  {C.~F.}}, \bibinfo {author} {\bibnamefont {Coroniti}, \bibfnamefont {F.~V.}},
  \bibinfo {author} {\bibnamefont {Kivelson}, \bibfnamefont {M.~G.}}, \bibinfo
  {author} {\bibnamefont {Pellat}, \bibfnamefont {R.}}, \bibinfo {author}
  {\bibnamefont {Walker}, \bibfnamefont {R.~J.}}, \bibinfo {author}
  {\bibnamefont {Luhr}, \bibfnamefont {H.}}, \ and\ \bibinfo {author}
  {\bibnamefont {Paschmann}, \bibfnamefont {G.}},\ }\href@noop {} {\bibfield
  {journal} {\bibinfo  {journal} {J. Geophys. Res.}\ }\textbf {\bibinfo
  {volume} {97}},\ \bibinfo {pages} {4027} (\bibinfo {year}
  {1992})}\BibitemShut {NoStop}%
\bibitem [{\citenamefont {{Aunai}}, \citenamefont {{Belmont}},\ and\
  \citenamefont {{Smets}}(2011)}]{Aunai11}%
  \BibitemOpen
  \bibfield  {author} {\bibinfo {author} {\bibnamefont {{Aunai}}, \bibfnamefont
  {N.}}, \bibinfo {author} {\bibnamefont {{Belmont}}, \bibfnamefont {G.}}, \
  and\ \bibinfo {author} {\bibnamefont {{Smets}}, \bibfnamefont {R.}},\ }\href
  {\doibase 10.1063/1.3664320} {\bibfield  {journal} {\bibinfo  {journal}
  {Phys. Plasmas}\ }\textbf {\bibinfo {volume} {18}},\ \bibinfo {eid} {122901}
  (\bibinfo {year} {2011})}\BibitemShut {NoStop}%
\bibitem [{\citenamefont {{Baker}}\ and\ \citenamefont
  {{Stone}}(1976)}]{Baker76}%
  \BibitemOpen
  \bibfield  {author} {\bibinfo {author} {\bibnamefont {{Baker}}, \bibfnamefont
  {D.~N.}}\ and\ \bibinfo {author} {\bibnamefont {{Stone}}, \bibfnamefont
  {E.~C.}},\ }\href {\doibase 10.1029/GL003i009p00557} {\bibfield  {journal}
  {\bibinfo  {journal} {Geophys. Res. Lett.}\ }\textbf {\bibinfo {volume}
  {3}},\ \bibinfo {pages} {557} (\bibinfo {year} {1976})}\BibitemShut {NoStop}%
\bibitem [{\citenamefont {{Birn}}\ \emph {et~al.}(2010)\citenamefont {{Birn}},
  \citenamefont {{Borovsky}}, \citenamefont {{Hesse}},\ and\ \citenamefont
  {{Schindler}}}]{Birn2010}%
  \BibitemOpen
  \bibfield  {author} {\bibinfo {author} {\bibnamefont {{Birn}}, \bibfnamefont
  {J.}}, \bibinfo {author} {\bibnamefont {{Borovsky}}, \bibfnamefont {J.~E.}},
  \bibinfo {author} {\bibnamefont {{Hesse}}, \bibfnamefont {M.}}, \ and\
  \bibinfo {author} {\bibnamefont {{Schindler}}, \bibfnamefont {K.}},\ }\href
  {\doibase 10.1063/1.3429676} {\bibfield  {journal} {\bibinfo  {journal}
  {Physics of Plasmas}\ }\textbf {\bibinfo {volume} {17}},\ \bibinfo {pages}
  {052108} (\bibinfo {year} {2010})}\BibitemShut {NoStop}%
\bibitem [{\citenamefont {Birn}\ \emph {et~al.}(2001)\citenamefont {Birn},
  \citenamefont {Drake}, \citenamefont {Shay}, \citenamefont {Rogers},
  \citenamefont {Denton}, \citenamefont {Hesse}, \citenamefont {Kuznetsova},
  \citenamefont {Ma}, \citenamefont {Bhattacharjee}, \citenamefont {Otto},\
  and\ \citenamefont {Pritchett}}]{Birn01}%
  \BibitemOpen
  \bibfield  {author} {\bibinfo {author} {\bibnamefont {Birn}, \bibfnamefont
  {J.}}, \bibinfo {author} {\bibnamefont {Drake}, \bibfnamefont {J.~F.}},
  \bibinfo {author} {\bibnamefont {Shay}, \bibfnamefont {M.~A.}}, \bibinfo
  {author} {\bibnamefont {Rogers}, \bibfnamefont {B.~N.}}, \bibinfo {author}
  {\bibnamefont {Denton}, \bibfnamefont {R.~E.}}, \bibinfo {author}
  {\bibnamefont {Hesse}, \bibfnamefont {M.}}, \bibinfo {author} {\bibnamefont
  {Kuznetsova}, \bibfnamefont {M.}}, \bibinfo {author} {\bibnamefont {Ma},
  \bibfnamefont {Z.~W.}}, \bibinfo {author} {\bibnamefont {Bhattacharjee},
  \bibfnamefont {A.}}, \bibinfo {author} {\bibnamefont {Otto}, \bibfnamefont
  {A.}}, \ and\ \bibinfo {author} {\bibnamefont {Pritchett}, \bibfnamefont
  {P.~L.}},\ }\href@noop {} {\bibfield  {journal} {\bibinfo  {journal} {J.
  Geophys. Res.}\ }\textbf {\bibinfo {volume} {106}},\ \bibinfo {pages} {3715}
  (\bibinfo {year} {2001})}\BibitemShut {NoStop}%
\bibitem [{\citenamefont {{Birn}}\ \emph {et~al.}(2000)\citenamefont {{Birn}},
  \citenamefont {{Thomsen}}, \citenamefont {{Borovsky}}, \citenamefont
  {{Reeves}},\ and\ \citenamefont {{Hesse}}}]{Birn00}%
  \BibitemOpen
  \bibfield  {author} {\bibinfo {author} {\bibnamefont {{Birn}}, \bibfnamefont
  {J.}}, \bibinfo {author} {\bibnamefont {{Thomsen}}, \bibfnamefont {M.~F.}},
  \bibinfo {author} {\bibnamefont {{Borovsky}}, \bibfnamefont {J.~E.}},
  \bibinfo {author} {\bibnamefont {{Reeves}}, \bibfnamefont {G.~D.}}, \ and\
  \bibinfo {author} {\bibnamefont {{Hesse}}, \bibfnamefont {M.}},\ }\href
  {\doibase 10.1063/1.874035} {\bibfield  {journal} {\bibinfo  {journal} {Phys.
  Plasmas}\ }\textbf {\bibinfo {volume} {7}},\ \bibinfo {pages} {2149}
  (\bibinfo {year} {2000})}\BibitemShut {NoStop}%
\bibitem [{\citenamefont {Cassak}\ and\ \citenamefont
  {Shay}(2007)}]{Cassak07d}%
  \BibitemOpen
  \bibfield  {author} {\bibinfo {author} {\bibnamefont {Cassak}, \bibfnamefont
  {P.~A.}}\ and\ \bibinfo {author} {\bibnamefont {Shay}, \bibfnamefont
  {M.~A.}},\ }\href@noop {} {\bibfield  {journal} {\bibinfo  {journal}
  {Phys.~Plasmas}\ }\textbf {\bibinfo {volume} {14}},\ \bibinfo {pages}
  {102114} (\bibinfo {year} {2007})}\BibitemShut {NoStop}%
\bibitem [{\citenamefont {Cattell}\ \emph {et~al.}(2005)\citenamefont
  {Cattell}, \citenamefont {Dombeck}, \citenamefont {Wygant}, \citenamefont
  {Drake}, \citenamefont {Swisdak}, \citenamefont {Goldstein}, \citenamefont
  {Keith}, \citenamefont {Fazakerley}, \citenamefont {Andr\'e}, \citenamefont
  {Lucek},\ and\ \citenamefont {Balogh}}]{Cattell05}%
  \BibitemOpen
  \bibfield  {author} {\bibinfo {author} {\bibnamefont {Cattell}, \bibfnamefont
  {C.}}, \bibinfo {author} {\bibnamefont {Dombeck}, \bibfnamefont {J.}},
  \bibinfo {author} {\bibnamefont {Wygant}, \bibfnamefont {J.}}, \bibinfo
  {author} {\bibnamefont {Drake}, \bibfnamefont {J.~F.}}, \bibinfo {author}
  {\bibnamefont {Swisdak}, \bibfnamefont {M.}}, \bibinfo {author} {\bibnamefont
  {Goldstein}, \bibfnamefont {M.~L.}}, \bibinfo {author} {\bibnamefont {Keith},
  \bibfnamefont {W.}}, \bibinfo {author} {\bibnamefont {Fazakerley},
  \bibfnamefont {A.}}, \bibinfo {author} {\bibnamefont {Andr\'e}, \bibfnamefont
  {M.}}, \bibinfo {author} {\bibnamefont {Lucek}, \bibfnamefont {E.}}, \ and\
  \bibinfo {author} {\bibnamefont {Balogh}, \bibfnamefont {A.}},\ }\href@noop
  {} {\bibfield  {journal} {\bibinfo  {journal} {J. Geophys. Res.}\ }\textbf
  {\bibinfo {volume} {110}},\ \bibinfo {pages} {A01211} (\bibinfo {year}
  {2005})}\BibitemShut {NoStop}%
\bibitem [{\citenamefont {Chen}\ \emph {et~al.}(2008)\citenamefont {Chen},
  \citenamefont {Bessho}, \citenamefont {Lefebvre}, \citenamefont {Vaith},
  \citenamefont {Fazakerley}, \citenamefont {Bhattacharjee}, \citenamefont
  {Puhl-Quinn}, \citenamefont {Runov}, \citenamefont {Khotyaintsev},
  \citenamefont {Vaivads} \emph {et~al.}}]{Chen08b}%
  \BibitemOpen
  \bibfield  {author} {\bibinfo {author} {\bibnamefont {Chen}, \bibfnamefont
  {L.}}, \bibinfo {author} {\bibnamefont {Bessho}, \bibfnamefont {N.}},
  \bibinfo {author} {\bibnamefont {Lefebvre}, \bibfnamefont {B.}}, \bibinfo
  {author} {\bibnamefont {Vaith}, \bibfnamefont {H.}}, \bibinfo {author}
  {\bibnamefont {Fazakerley}, \bibfnamefont {A.}}, \bibinfo {author}
  {\bibnamefont {Bhattacharjee}, \bibfnamefont {A.}}, \bibinfo {author}
  {\bibnamefont {Puhl-Quinn}, \bibfnamefont {P.}}, \bibinfo {author}
  {\bibnamefont {Runov}, \bibfnamefont {A.}}, \bibinfo {author} {\bibnamefont
  {Khotyaintsev}, \bibfnamefont {Y.}}, \bibinfo {author} {\bibnamefont
  {Vaivads}, \bibfnamefont {A.}},  \emph {et~al.},\ }\href@noop {} {\bibfield
  {journal} {\bibinfo  {journal} {J. Geophys. Res}\ }\textbf {\bibinfo {volume}
  {113}},\ \bibinfo {pages} {A12213} (\bibinfo {year} {2008})}\BibitemShut
  {NoStop}%
\bibitem [{\citenamefont {{Chen}}\ \emph {et~al.}(2008)\citenamefont {{Chen}},
  \citenamefont {{Bhattacharjee}}, \citenamefont {{Puhl-Quinn}}, \citenamefont
  {{Yang}}, \citenamefont {{Bessho}}, \citenamefont {{Imada}}, \citenamefont
  {{M{\"u}hlbachler}}, \citenamefont {{Daly}}, \citenamefont {{Lefebvre}},
  \citenamefont {{Khotyaintsev}}, \citenamefont {{Vaivads}}, \citenamefont
  {{Fazakerley}},\ and\ \citenamefont {{Georgescu}}}]{Chen08}%
  \BibitemOpen
  \bibfield  {author} {\bibinfo {author} {\bibnamefont {{Chen}}, \bibfnamefont
  {L.}}, \bibinfo {author} {\bibnamefont {{Bhattacharjee}}, \bibfnamefont
  {A.}}, \bibinfo {author} {\bibnamefont {{Puhl-Quinn}}, \bibfnamefont
  {P.~A.}}, \bibinfo {author} {\bibnamefont {{Yang}}, \bibfnamefont {H.}},
  \bibinfo {author} {\bibnamefont {{Bessho}}, \bibfnamefont {N.}}, \bibinfo
  {author} {\bibnamefont {{Imada}}, \bibfnamefont {S.}}, \bibinfo {author}
  {\bibnamefont {{M{\"u}hlbachler}}, \bibfnamefont {S.}}, \bibinfo {author}
  {\bibnamefont {{Daly}}, \bibfnamefont {P.~W.}}, \bibinfo {author}
  {\bibnamefont {{Lefebvre}}, \bibfnamefont {B.}}, \bibinfo {author}
  {\bibnamefont {{Khotyaintsev}}, \bibfnamefont {Y.}}, \bibinfo {author}
  {\bibnamefont {{Vaivads}}, \bibfnamefont {A.}}, \bibinfo {author}
  {\bibnamefont {{Fazakerley}}, \bibfnamefont {A.}}, \ and\ \bibinfo {author}
  {\bibnamefont {{Georgescu}}, \bibfnamefont {E.}},\ }\href {\doibase
  10.1038/nphys777} {\bibfield  {journal} {\bibinfo  {journal} {Nature
  Physics}\ }\textbf {\bibinfo {volume} {4}},\ \bibinfo {pages} {19} (\bibinfo
  {year} {2008})}\BibitemShut {NoStop}%
\bibitem [{\citenamefont {{Dahlin}}, \citenamefont {{Drake}},\ and\
  \citenamefont {{Swisdak}}(2014)}]{Dahlin14}%
  \BibitemOpen
  \bibfield  {author} {\bibinfo {author} {\bibnamefont {{Dahlin}},
  \bibfnamefont {J.~T.}}, \bibinfo {author} {\bibnamefont {{Drake}},
  \bibfnamefont {J.~F.}}, \ and\ \bibinfo {author} {\bibnamefont {{Swisdak}},
  \bibfnamefont {M.}},\ }\href {\doibase doi: 10.1063/1.4894484} {\bibfield
  {journal} {\bibinfo  {journal} {Phys. Plasmas}\ }\textbf {\bibinfo {volume}
  {21}},\ \bibinfo {pages} {092304} (\bibinfo {year} {2014})}\BibitemShut
  {NoStop}%
\bibitem [{\citenamefont {Drake}\ \emph {et~al.}(2006)\citenamefont {Drake},
  \citenamefont {Swisdak}, \citenamefont {Che},\ and\ \citenamefont
  {Shay}}]{Drake06b}%
  \BibitemOpen
  \bibfield  {author} {\bibinfo {author} {\bibnamefont {Drake}, \bibfnamefont
  {J.~F.}}, \bibinfo {author} {\bibnamefont {Swisdak}, \bibfnamefont {M.}},
  \bibinfo {author} {\bibnamefont {Che}, \bibfnamefont {H.}}, \ and\ \bibinfo
  {author} {\bibnamefont {Shay}, \bibfnamefont {M.~A.}},\ }\href@noop {}
  {\bibfield  {journal} {\bibinfo  {journal} {Nature}\ }\textbf {\bibinfo
  {volume} {443}},\ \bibinfo {pages} {553} (\bibinfo {year}
  {2006})}\BibitemShut {NoStop}%
\bibitem [{\citenamefont {{Eastwood}}\ \emph {et~al.}(2013)\citenamefont
  {{Eastwood}}, \citenamefont {{Phan}}, \citenamefont {{Drake}}, \citenamefont
  {{Shay}}, \citenamefont {{Borg}}, \citenamefont {{Lavraud}},\ and\
  \citenamefont {{Taylor}}}]{Eastwood13}%
  \BibitemOpen
  \bibfield  {author} {\bibinfo {author} {\bibnamefont {{Eastwood}},
  \bibfnamefont {J.~P.}}, \bibinfo {author} {\bibnamefont {{Phan}},
  \bibfnamefont {T.~D.}}, \bibinfo {author} {\bibnamefont {{Drake}},
  \bibfnamefont {J.~F.}}, \bibinfo {author} {\bibnamefont {{Shay}},
  \bibfnamefont {M.~A.}}, \bibinfo {author} {\bibnamefont {{Borg}},
  \bibfnamefont {A.~L.}}, \bibinfo {author} {\bibnamefont {{Lavraud}},
  \bibfnamefont {B.}}, \ and\ \bibinfo {author} {\bibnamefont {{Taylor}},
  \bibfnamefont {M.~G.~G.~T.}},\ }\href {\doibase
  10.1103/PhysRevLett.110.225001} {\bibfield  {journal} {\bibinfo  {journal}
  {Phys. Rev. Lett.}\ }\textbf {\bibinfo {volume} {110}},\ \bibinfo {eid}
  {225001} (\bibinfo {year} {2013})}\BibitemShut {NoStop}%
\bibitem [{\citenamefont {{Egedal}}, \citenamefont {{Daughton}},\ and\
  \citenamefont {{Le}}(2012)}]{Egedal12}%
  \BibitemOpen
  \bibfield  {author} {\bibinfo {author} {\bibnamefont {{Egedal}},
  \bibfnamefont {J.}}, \bibinfo {author} {\bibnamefont {{Daughton}},
  \bibfnamefont {W.}}, \ and\ \bibinfo {author} {\bibnamefont {{Le}},
  \bibfnamefont {A.}},\ }\href {\doibase 10.1038/nphys2249} {\bibfield
  {journal} {\bibinfo  {journal} {Nature Physics}\ }\textbf {\bibinfo {volume}
  {8}},\ \bibinfo {pages} {321} (\bibinfo {year} {2012})}\BibitemShut {NoStop}%
\bibitem [{\citenamefont {{Egedal}}, \citenamefont {{Le}},\ and\ \citenamefont
  {{Daughton}}(2013)}]{Egedal13}%
  \BibitemOpen
  \bibfield  {author} {\bibinfo {author} {\bibnamefont {{Egedal}},
  \bibfnamefont {J.}}, \bibinfo {author} {\bibnamefont {{Le}}, \bibfnamefont
  {A.}}, \ and\ \bibinfo {author} {\bibnamefont {{Daughton}}, \bibfnamefont
  {W.}},\ }\href {\doibase 10.1063/1.4811092} {\bibfield  {journal} {\bibinfo
  {journal} {Phys. Plasmas}\ }\textbf {\bibinfo {volume} {20}},\ \bibinfo
  {pages} {061201} (\bibinfo {year} {2013})}\BibitemShut {NoStop}%
\bibitem [{\citenamefont {Egedal}\ \emph {et~al.}(2010)\citenamefont {Egedal},
  \citenamefont {L{\^e}}, \citenamefont {Zhu}, \citenamefont {Daughton},
  \citenamefont {{\O}ieroset}, \citenamefont {Phan}, \citenamefont {Lin},\ and\
  \citenamefont {Eastwood}}]{Egedal10}%
  \BibitemOpen
  \bibfield  {author} {\bibinfo {author} {\bibnamefont {Egedal}, \bibfnamefont
  {J.}}, \bibinfo {author} {\bibnamefont {L{\^e}}, \bibfnamefont {A.}},
  \bibinfo {author} {\bibnamefont {Zhu}, \bibfnamefont {Y.}}, \bibinfo {author}
  {\bibnamefont {Daughton}, \bibfnamefont {W.}}, \bibinfo {author}
  {\bibnamefont {{\O}ieroset}, \bibfnamefont {M.}}, \bibinfo {author}
  {\bibnamefont {Phan}, \bibfnamefont {T.}}, \bibinfo {author} {\bibnamefont
  {Lin}, \bibfnamefont {R.}}, \ and\ \bibinfo {author} {\bibnamefont
  {Eastwood}, \bibfnamefont {J.}},\ }\href@noop {} {\bibfield  {journal}
  {\bibinfo  {journal} {Geophys. Res. Lett}\ }\textbf {\bibinfo {volume}
  {37}},\ \bibinfo {pages} {L10102} (\bibinfo {year} {2010})}\BibitemShut
  {NoStop}%
\bibitem [{\citenamefont {Fujimoto}\ and\ \citenamefont
  {Machida}(2006)}]{Fujimoto06b}%
  \BibitemOpen
  \bibfield  {author} {\bibinfo {author} {\bibnamefont {Fujimoto},
  \bibfnamefont {K.}}\ and\ \bibinfo {author} {\bibnamefont {Machida},
  \bibfnamefont {S.}},\ }\href@noop {} {\bibfield  {journal} {\bibinfo
  {journal} {Journal of geophysical research}\ }\textbf {\bibinfo {volume}
  {111}},\ \bibinfo {pages} {A09216} (\bibinfo {year} {2006})}\BibitemShut
  {NoStop}%
\bibitem [{\citenamefont {{Gosling}}\ \emph {et~al.}(2007)\citenamefont
  {{Gosling}}, \citenamefont {{Eriksson}}, \citenamefont {{Phan}},
  \citenamefont {{Larson}}, \citenamefont {{Skoug}},\ and\ \citenamefont
  {{McComas}}}]{Gosling07}%
  \BibitemOpen
  \bibfield  {author} {\bibinfo {author} {\bibnamefont {{Gosling}},
  \bibfnamefont {J.~T.}}, \bibinfo {author} {\bibnamefont {{Eriksson}},
  \bibfnamefont {S.}}, \bibinfo {author} {\bibnamefont {{Phan}}, \bibfnamefont
  {T.~D.}}, \bibinfo {author} {\bibnamefont {{Larson}}, \bibfnamefont {D.~E.}},
  \bibinfo {author} {\bibnamefont {{Skoug}}, \bibfnamefont {R.~M.}}, \ and\
  \bibinfo {author} {\bibnamefont {{McComas}}, \bibfnamefont {D.~J.}},\ }\href
  {\doibase 10.1029/2006GL029033} {\bibfield  {journal} {\bibinfo  {journal}
  {Geophys. Res. Lett.}\ }\textbf {\bibinfo {volume} {34}},\ \bibinfo {eid}
  {L06102} (\bibinfo {year} {2007})}\BibitemShut {NoStop}%
\bibitem [{\citenamefont {Gosling}\ \emph {et~al.}(2006)\citenamefont
  {Gosling}, \citenamefont {Eriksson}, \citenamefont {Skoug}, \citenamefont
  {McComas},\ and\ \citenamefont {Forsyth}}]{Gosling06}%
  \BibitemOpen
  \bibfield  {author} {\bibinfo {author} {\bibnamefont {Gosling}, \bibfnamefont
  {J.~T.}}, \bibinfo {author} {\bibnamefont {Eriksson}, \bibfnamefont {S.}},
  \bibinfo {author} {\bibnamefont {Skoug}, \bibfnamefont {R.~M.}}, \bibinfo
  {author} {\bibnamefont {McComas}, \bibfnamefont {D.~J.}}, \ and\ \bibinfo
  {author} {\bibnamefont {Forsyth}, \bibfnamefont {R.~J.}},\ }\href@noop {}
  {\bibfield  {journal} {\bibinfo  {journal} {Ap.~J.}\ }\textbf {\bibinfo
  {volume} {644}},\ \bibinfo {pages} {613} (\bibinfo {year}
  {2006})}\BibitemShut {NoStop}%
\bibitem [{\citenamefont {Horiuchi}\ and\ \citenamefont
  {Sato}(1997)}]{Horiuchi97}%
  \BibitemOpen
  \bibfield  {author} {\bibinfo {author} {\bibnamefont {Horiuchi},
  \bibfnamefont {R.}}\ and\ \bibinfo {author} {\bibnamefont {Sato},
  \bibfnamefont {T.}},\ }\href@noop {} {\bibfield  {journal} {\bibinfo
  {journal} {Phys. Plasmas}\ }\textbf {\bibinfo {volume} {4}},\ \bibinfo
  {pages} {277} (\bibinfo {year} {1997})}\BibitemShut {NoStop}%
\bibitem [{\citenamefont {Hoshino}, \citenamefont {Hiraide},\ and\
  \citenamefont {Mukai}(2001)}]{Hoshino01a}%
  \BibitemOpen
  \bibfield  {author} {\bibinfo {author} {\bibnamefont {Hoshino}, \bibfnamefont
  {M.}}, \bibinfo {author} {\bibnamefont {Hiraide}, \bibfnamefont {K.}}, \ and\
  \bibinfo {author} {\bibnamefont {Mukai}, \bibfnamefont {T.}},\ }\href@noop {}
  {\bibfield  {journal} {\bibinfo  {journal} {Earth Planets Space}\ }\textbf
  {\bibinfo {volume} {53}},\ \bibinfo {pages} {627} (\bibinfo {year}
  {2001})}\BibitemShut {NoStop}%
\bibitem [{\citenamefont {Hoshino}\ \emph {et~al.}(2001)\citenamefont
  {Hoshino}, \citenamefont {Mukai}, \citenamefont {Terasawa},\ and\
  \citenamefont {Shinohara}}]{Hoshino01b}%
  \BibitemOpen
  \bibfield  {author} {\bibinfo {author} {\bibnamefont {Hoshino}, \bibfnamefont
  {M.}}, \bibinfo {author} {\bibnamefont {Mukai}, \bibfnamefont {T.}}, \bibinfo
  {author} {\bibnamefont {Terasawa}, \bibfnamefont {T.}}, \ and\ \bibinfo
  {author} {\bibnamefont {Shinohara}, \bibfnamefont {I.}},\ }\href@noop {}
  {\bibfield  {journal} {\bibinfo  {journal} {J. Geophys. Res.}\ }\textbf
  {\bibinfo {volume} {106}},\ \bibinfo {pages} {25979} (\bibinfo {year}
  {2001})}\BibitemShut {NoStop}%
\bibitem [{\citenamefont {Karimabadi}, \citenamefont {Daughton},\ and\
  \citenamefont {Scudder}(2007)}]{Karimabadi07}%
  \BibitemOpen
  \bibfield  {author} {\bibinfo {author} {\bibnamefont {Karimabadi},
  \bibfnamefont {H.}}, \bibinfo {author} {\bibnamefont {Daughton},
  \bibfnamefont {W.}}, \ and\ \bibinfo {author} {\bibnamefont {Scudder},
  \bibfnamefont {J.}},\ }\href@noop {} {\bibfield  {journal} {\bibinfo
  {journal} {Geophys. Res. Lett.}\ }\textbf {\bibinfo {volume} {34}},\ \bibinfo
  {pages} {L13104, doi:10.1029/2007GL030306} (\bibinfo {year}
  {2007})}\BibitemShut {NoStop}%
\bibitem [{\citenamefont {Le}\ \emph {et~al.}(2009)\citenamefont {Le},
  \citenamefont {Egedal}, \citenamefont {Daughton}, \citenamefont {Fox},\ and\
  \citenamefont {Katz}}]{Le09}%
  \BibitemOpen
  \bibfield  {author} {\bibinfo {author} {\bibnamefont {Le}, \bibfnamefont
  {A.}}, \bibinfo {author} {\bibnamefont {Egedal}, \bibfnamefont {J.}},
  \bibinfo {author} {\bibnamefont {Daughton}, \bibfnamefont {W.}}, \bibinfo
  {author} {\bibnamefont {Fox}, \bibfnamefont {W.}}, \ and\ \bibinfo {author}
  {\bibnamefont {Katz}, \bibfnamefont {N.}},\ }\href {\doibase
  10.1103/PhysRevLett.102.085001} {\bibfield  {journal} {\bibinfo  {journal}
  {Phys. Rev. Lett.}\ }\textbf {\bibinfo {volume} {102}},\ \bibinfo {pages}
  {085001} (\bibinfo {year} {2009})}\BibitemShut {NoStop}%
\bibitem [{\citenamefont {Oieroset}\ \emph {et~al.}(2002)\citenamefont
  {Oieroset}, \citenamefont {Lin}, \citenamefont {Phan}, \citenamefont
  {Larson},\ and\ \citenamefont {Bale}}]{Oieroset02}%
  \BibitemOpen
  \bibfield  {author} {\bibinfo {author} {\bibnamefont {Oieroset},
  \bibfnamefont {M.}}, \bibinfo {author} {\bibnamefont {Lin}, \bibfnamefont
  {R.~P.}}, \bibinfo {author} {\bibnamefont {Phan}, \bibfnamefont {T.~D.}},
  \bibinfo {author} {\bibnamefont {Larson}, \bibfnamefont {D.~E.}}, \ and\
  \bibinfo {author} {\bibnamefont {Bale}, \bibfnamefont {S.~D.}},\ }\href@noop
  {} {\bibfield  {journal} {\bibinfo  {journal} {Phys. Rev. Lett.}\ }\textbf
  {\bibinfo {volume} {89}},\ \bibinfo {pages} {195001} (\bibinfo {year}
  {2002})}\BibitemShut {NoStop}%
\bibitem [{\citenamefont {{Phan}}\ \emph {et~al.}(2013)\citenamefont {{Phan}},
  \citenamefont {{Shay}}, \citenamefont {{Gosling}}, \citenamefont
  {{Fujimoto}}, \citenamefont {{Drake}}, \citenamefont {{Paschmann}},
  \citenamefont {{Oieroset}}, \citenamefont {{Eastwood}},\ and\ \citenamefont
  {{Angelopoulos}}}]{Phan13}%
  \BibitemOpen
  \bibfield  {author} {\bibinfo {author} {\bibnamefont {{Phan}}, \bibfnamefont
  {T.~D.}}, \bibinfo {author} {\bibnamefont {{Shay}}, \bibfnamefont {M.~A.}},
  \bibinfo {author} {\bibnamefont {{Gosling}}, \bibfnamefont {J.~T.}}, \bibinfo
  {author} {\bibnamefont {{Fujimoto}}, \bibfnamefont {M.}}, \bibinfo {author}
  {\bibnamefont {{Drake}}, \bibfnamefont {J.~F.}}, \bibinfo {author}
  {\bibnamefont {{Paschmann}}, \bibfnamefont {G.}}, \bibinfo {author}
  {\bibnamefont {{Oieroset}}, \bibfnamefont {M.}}, \bibinfo {author}
  {\bibnamefont {{Eastwood}}, \bibfnamefont {J.~P.}}, \ and\ \bibinfo {author}
  {\bibnamefont {{Angelopoulos}}, \bibfnamefont {V.}},\ }\href {\doibase
  10.1002/grl.50917} {\bibfield  {journal} {\bibinfo  {journal} {Geophys. Res.
  Lett.}\ }\textbf {\bibinfo {volume} {40}},\ \bibinfo {pages} {4475} (\bibinfo
  {year} {2013})}\BibitemShut {NoStop}%
\bibitem [{\citenamefont {Priest}\ and\ \citenamefont
  {Forbes}(2000)}]{Priest00}%
  \BibitemOpen
  \bibfield  {author} {\bibinfo {author} {\bibnamefont {Priest}, \bibfnamefont
  {E.}}\ and\ \bibinfo {author} {\bibnamefont {Forbes}, \bibfnamefont {T.}},\
  }\href@noop {} {\emph {\bibinfo {title} {Magnetic Reconnection, {MHD} Theory
  and Applications}}}\ (\bibinfo  {publisher} {Cambridge University Press},\
  \bibinfo {address} {New York, NY},\ \bibinfo {year} {2000})\BibitemShut
  {NoStop}%
\bibitem [{\citenamefont {Pritchett}(2006)}]{Pritchett06}%
  \BibitemOpen
  \bibfield  {author} {\bibinfo {author} {\bibnamefont {Pritchett},
  \bibfnamefont {P.~L.}},\ }\href@noop {} {\bibfield  {journal} {\bibinfo
  {journal} {Geophys. Res. Lett.}\ }\textbf {\bibinfo {volume} {33}},\ \bibinfo
  {pages} {L13104} (\bibinfo {year} {2006})},\ \bibinfo {note}
  {doi:10.1029/2005GL025267}\BibitemShut {NoStop}%
\bibitem [{\citenamefont {{Pulupa}}\ \emph {et~al.}(2014)\citenamefont
  {{Pulupa}}, \citenamefont {{Salem}}, \citenamefont {{Phan}}, \citenamefont
  {{Gosling}},\ and\ \citenamefont {{Bale}}}]{Pulupa14}%
  \BibitemOpen
  \bibfield  {author} {\bibinfo {author} {\bibnamefont {{Pulupa}},
  \bibfnamefont {M.~P.}}, \bibinfo {author} {\bibnamefont {{Salem}},
  \bibfnamefont {C.}}, \bibinfo {author} {\bibnamefont {{Phan}}, \bibfnamefont
  {T.~D.}}, \bibinfo {author} {\bibnamefont {{Gosling}}, \bibfnamefont
  {J.~T.}}, \ and\ \bibinfo {author} {\bibnamefont {{Bale}}, \bibfnamefont
  {S.~D.}},\ }\href {\doibase 10.1088/2041-8205/791/1/L17} {\bibfield
  {journal} {\bibinfo  {journal} {Astrophys. J. Lett.}\ }\textbf {\bibinfo
  {volume} {791}},\ \bibinfo {eid} {L17} (\bibinfo {year} {2014})}\BibitemShut
  {NoStop}%
\bibitem [{\citenamefont {{Retin{\`o}}}\ \emph {et~al.}(2008)\citenamefont
  {{Retin{\`o}}}, \citenamefont {{Nakamura}}, \citenamefont {{Vaivads}},
  \citenamefont {{Khotyaintsev}}, \citenamefont {{Hayakawa}}, \citenamefont
  {{Tanaka}}, \citenamefont {{Kasahara}}, \citenamefont {{Fujimoto}},
  \citenamefont {{Shinohara}}, \citenamefont {{Eastwood}}, \citenamefont
  {{Andr{\'e}}}, \citenamefont {{Baumjohann}}, \citenamefont {{Daly}},
  \citenamefont {{Kronberg}},\ and\ \citenamefont
  {{Cornilleau-Wehrlin}}}]{Retino08}%
  \BibitemOpen
  \bibfield  {author} {\bibinfo {author} {\bibnamefont {{Retin{\`o}}},
  \bibfnamefont {A.}}, \bibinfo {author} {\bibnamefont {{Nakamura}},
  \bibfnamefont {R.}}, \bibinfo {author} {\bibnamefont {{Vaivads}},
  \bibfnamefont {A.}}, \bibinfo {author} {\bibnamefont {{Khotyaintsev}},
  \bibfnamefont {Y.}}, \bibinfo {author} {\bibnamefont {{Hayakawa}},
  \bibfnamefont {T.}}, \bibinfo {author} {\bibnamefont {{Tanaka}},
  \bibfnamefont {K.}}, \bibinfo {author} {\bibnamefont {{Kasahara}},
  \bibfnamefont {S.}}, \bibinfo {author} {\bibnamefont {{Fujimoto}},
  \bibfnamefont {M.}}, \bibinfo {author} {\bibnamefont {{Shinohara}},
  \bibfnamefont {I.}}, \bibinfo {author} {\bibnamefont {{Eastwood}},
  \bibfnamefont {J.~P.}}, \bibinfo {author} {\bibnamefont {{Andr{\'e}}},
  \bibfnamefont {M.}}, \bibinfo {author} {\bibnamefont {{Baumjohann}},
  \bibfnamefont {W.}}, \bibinfo {author} {\bibnamefont {{Daly}}, \bibfnamefont
  {P.~W.}}, \bibinfo {author} {\bibnamefont {{Kronberg}}, \bibfnamefont
  {E.~A.}}, \ and\ \bibinfo {author} {\bibnamefont {{Cornilleau-Wehrlin}},
  \bibfnamefont {N.}},\ }\href {\doibase 10.1029/2008JA013511} {\bibfield
  {journal} {\bibinfo  {journal} {J. Geophys. Res.}\ }\textbf {\bibinfo
  {volume} {113}},\ \bibinfo {eid} {A12215} (\bibinfo {year}
  {2008})}\BibitemShut {NoStop}%
\bibitem [{\citenamefont {Shay}, \citenamefont {Drake},\ and\ \citenamefont
  {Swisdak}(2007)}]{Shay07}%
  \BibitemOpen
  \bibfield  {author} {\bibinfo {author} {\bibnamefont {Shay}, \bibfnamefont
  {M.~A.}}, \bibinfo {author} {\bibnamefont {Drake}, \bibfnamefont {J.~F.}}, \
  and\ \bibinfo {author} {\bibnamefont {Swisdak}, \bibfnamefont {M.}},\
  }\href@noop {} {\bibfield  {journal} {\bibinfo  {journal} {Phys. Rev. Lett.}\
  }\textbf {\bibinfo {volume} {99}},\ \bibinfo {pages} {155002} (\bibinfo
  {year} {2007})}\BibitemShut {NoStop}%
\bibitem [{\citenamefont {Swisdak}\ and\ \citenamefont
  {Drake}(2007)}]{Swisdak07}%
  \BibitemOpen
  \bibfield  {author} {\bibinfo {author} {\bibnamefont {Swisdak}, \bibfnamefont
  {M.}}\ and\ \bibinfo {author} {\bibnamefont {Drake}, \bibfnamefont {J.~F.}},\
  }\href@noop {} {\bibfield  {journal} {\bibinfo  {journal} {Geophys. Res.
  Lett.}\ }\textbf {\bibinfo {volume} {34}},\ \bibinfo {pages} {L11106.
  doi:10.1029/2007GL029815} (\bibinfo {year} {2007})}\BibitemShut {NoStop}%
\bibitem [{\citenamefont {Swisdak}\ \emph {et~al.}(2005)\citenamefont
  {Swisdak}, \citenamefont {Drake}, \citenamefont {McIlhargey},\ and\
  \citenamefont {Shay}}]{Swisdak05}%
  \BibitemOpen
  \bibfield  {author} {\bibinfo {author} {\bibnamefont {Swisdak}, \bibfnamefont
  {M.}}, \bibinfo {author} {\bibnamefont {Drake}, \bibfnamefont {J.~F.}},
  \bibinfo {author} {\bibnamefont {McIlhargey}, \bibfnamefont {J.~G.}}, \ and\
  \bibinfo {author} {\bibnamefont {Shay}, \bibfnamefont {M.~A.}},\ }\href@noop
  {} {\bibfield  {journal} {\bibinfo  {journal} {J. Geophys. Res.}\ }\textbf
  {\bibinfo {volume} {110}},\ \bibinfo {pages} {A05210} (\bibinfo {year}
  {2005})}\BibitemShut {NoStop}%
\bibitem [{\citenamefont {Yamada}\ \emph {et~al.}(2014)\citenamefont {Yamada},
  \citenamefont {Yoo}, \citenamefont {Jara-Almonte}, \citenamefont {Ji},
  \citenamefont {Kulsrud},\ and\ \citenamefont {Myers}}]{Yamada14}%
  \BibitemOpen
  \bibfield  {author} {\bibinfo {author} {\bibnamefont {Yamada}, \bibfnamefont
  {M.}}, \bibinfo {author} {\bibnamefont {Yoo}, \bibfnamefont {J.}}, \bibinfo
  {author} {\bibnamefont {Jara-Almonte}, \bibfnamefont {J.}}, \bibinfo {author}
  {\bibnamefont {Ji}, \bibfnamefont {H.}}, \bibinfo {author} {\bibnamefont
  {Kulsrud}, \bibfnamefont {R.~M.}}, \ and\ \bibinfo {author} {\bibnamefont
  {Myers}, \bibfnamefont {C.~E.}},\ }\href {\doibase doi:10.1038/ncomms5774}
  {\bibfield  {journal} {\bibinfo  {journal} {Nature Communications}\ }\textbf
  {\bibinfo {volume} {5}},\ \bibinfo {pages} {4774} (\bibinfo {year}
  {2014})}\BibitemShut {NoStop}%
\bibitem [{\citenamefont {Zeiler}\ \emph {et~al.}(2002)\citenamefont {Zeiler},
  \citenamefont {Biskamp}, \citenamefont {Drake}, \citenamefont {Rogers},
  \citenamefont {Shay},\ and\ \citenamefont {Scholer}}]{Zeiler02}%
  \BibitemOpen
  \bibfield  {author} {\bibinfo {author} {\bibnamefont {Zeiler}, \bibfnamefont
  {A.}}, \bibinfo {author} {\bibnamefont {Biskamp}, \bibfnamefont {D.}},
  \bibinfo {author} {\bibnamefont {Drake}, \bibfnamefont {J.~F.}}, \bibinfo
  {author} {\bibnamefont {Rogers}, \bibfnamefont {B.~N.}}, \bibinfo {author}
  {\bibnamefont {Shay}, \bibfnamefont {M.~A.}}, \ and\ \bibinfo {author}
  {\bibnamefont {Scholer}, \bibfnamefont {M.}},\ }\href@noop {} {\bibfield
  {journal} {\bibinfo  {journal} {J. Geophys. Res.}\ }\textbf {\bibinfo
  {volume} {107}},\ \bibinfo {pages} {1230} (\bibinfo {year} {2002})},\
  \bibinfo {note} {doi:10.1029/2001JA000287}\BibitemShut {NoStop}%
\end{thebibliography}%

\newpage

\begin{table}[h]
\begin{center} \tiny
\begin{tabular}{|c|c|c|c|c|c|c|c|c|}\hline
Run & \parbox{.5in}{\(m_{i}/m_{e}\) compare} & \(m_{i}/m_{e}\) & \(B_{r}\) & \(B_{g}\) &  \(n_{in}\)& \(T_{e}\) & \(T_i\) &
\parbox{.5in}{Reference Number}\ \\\hline
1 &  & 25 & 1.000 & 0.000 & 0.20 & 0.250 & 0.250 & 301 \\\hline
2 &  & 25 & 1.000 & 1.000 & 0.20 & 0.250 & 0.250 & 302 \\\hline
3 &  & 25 & 1.000 & 0.000 & 0.20 & 0.250 & 2.250 & 303 \\\hline
4 &  & 25 & 1.000 & 1.000 & 0.20 & 0.250 & 2.250 & 304 \\\hline
5 &  & 25 & 1.000 & 0.000 & 1.00 & 0.250 & 0.250 & 307 \\\hline
6 &  & 25 & 1.000 & 1.000 & 1.00 & 0.250 & 0.250 & 311 \\\hline
7 &  & 25 & 0.447 & 0.000 & 0.20 & 0.250 & 0.250 & 308001 \\\hline
8 &  & 25 & 0.447 & 0.447 & 0.20 & 0.250 & 0.250 & 312001 \\\hline
9 &  & 25 & 1.000 & 0.000 & 0.04 & 0.250 & 2.250 & 309 \\\hline
10 &  & 25 & 1.000 & 1.000 & 0.04 & 0.250 & 2.250 & 313 \\\hline
11 &  & 25 & 2.236 & 0.000 & 0.20 & 0.250 & 2.250 & 310001 \\\hline
12 &  & 25 & 2.236 & 2.236 & 0.20 & 0.250 & 2.250 & 314001 \\\hline
13 &  & 25 & 0.447 & 0.000 & 0.20 & 0.250 & 2.250 & 319 \\\hline
14 &  & 25 & 0.447 & 0.447 & 0.20 & 0.250 & 2.250 & 320 \\\hline
15 &  & 25 & 1.000 & 0.000 & 1.00 & 0.250 & 2.250 & 321 \\\hline
16 &  & 25 & 1.000 & 1.000 & 1.00 & 0.250 & 2.250 & 322 \\\hline
17 &  & 25 & 1.000 & 0.000 & 0.20 & 0.250 & 1.250 & 323 \\\hline
18 &  & 25 & 1.000 & 1.000 & 0.20 & 0.250 & 1.250 & 324 \\\hline
19 & \(\checkmark\) & 25 & 1.000 & 0.000 & 0.20 & 0.063 & 0.313 & 325 \\\hline
20 & \(\checkmark\) & 25 & 1.000 & 1.000 & 0.20 & 0.063 & 0.313 & 326 \\\hline
21 &  & 25 & 1.000 & 1.000 & 0.20 & 1.000 & 5.000 & 328 \\\hline
22 & \(\checkmark\) & 25 & 1.000 & 0.000 & 0.20 & 0.250 & 1.250 & 601 \\\hline
23 & \(\checkmark\) & 25 & 1.000 & 1.000 & 0.20 & 0.250 & 1.250 & 604 \\\hline
24 & \(\checkmark\) & 25 & 0.447 & 0.000 & 0.20 & 0.250 & 1.250 & 602 \\\hline
25 & \(\checkmark\) & 25 & 2.236 & 0.000 & 0.20 & 0.250 & 1.250 & 603 \\\hline
26 &  & 25 & 1.000 & 0.000 & 0.20 & 0.250 & 1.250 & 621 \\\hline
27 &  & 25 & 0.447 & 0.000 & 0.20 & 0.250 & 1.250 & 622 \\\hline
28 &  & 25 & 2.236 & 0.000 & 0.20 & 0.250 & 1.250 & 623 \\\hline
29 &  & 25 & 1.000 & 1.000 & 0.20 & 0.250 & 1.250 & 624 \\\hline
30 & \(\checkmark\) & 25 & 0.447 & 0.447 & 0.20 & 0.250 & 1.250 & 625 \\\hline
31 &  & 25 & 2.236 & 2.236 & 0.20 & 0.250 & 1.250 & 626 \\\hline
32 &  & 25 & 1.000 & 0.000 & 0.20 & 1.000 & 1.000 & 641 \\\hline
33 & \(\checkmark\) & 25 & 2.236 & 0.000 & 0.20 & 1.250 & 6.250 & 651 \\\hline
34 &  & 25 & 0.447 & 0.000 & 0.20 & 0.050 & 0.250 & 652 \\\hline
35 &  & 25 & 1.000 & 0.000 & 0.04 & 1.250 & 6.250 & 655 \\\hline
36 &  & 25 & 0.447 & 0.000 & 0.04 & 0.250 & 1.250 & 657 \\\hline
37 & \(\checkmark\) & 25 & 1.673 & 0.000 & 0.20 & 0.700 & 3.500 & 661 \\\hline
38 & \(\checkmark\) & 25 & 0.748 & 0.000 & 0.04 & 0.700 & 3.500 & 662 \\\hline
39 &  & 25 & 1.000 & 0.000 & 0.20 & 0.750 & 0.750 & 671 \\\hline
40 &  & 25 & 0.447 & 0.000 & 0.20 & 0.150 & 0.150 & 672 \\\hline
41 &  & 25 & 1.000 & 0.000 & 0.20 & 0.150 & 1.350 & 674 \\\hline
42 &  & 25 & 0.447 & 0.000 & 0.20 & 0.030 & 0.270 & 675 \\\hline
43 &  & 25 & 2.236 & 0.000 & 0.20 & 0.750 & 6.750 & 676 \\\hline
44 &  & 25 & 0.447 & 0.447 & 0.20 & 0.050 & 0.250 & 681 \\\hline
45 &  & 25 & 2.236 & 2.236 & 0.20 & 1.250 & 6.250 & 682 \\\hline
46 & \(\checkmark\) & 100 & 1.000 & 0.000 & 0.20 & 0.250 & 1.250 & 701 \\\hline
47 & \(\checkmark\) & 100 & 1.000 & 1.000 & 0.20 & 0.250 & 1.250 & 702 \\\hline
48 & \(\checkmark\) & 100 & 0.447 & 0.000 & 0.20 & 0.250 & 1.250 & 703 \\\hline
49 & \(\checkmark\) & 100 & 0.447 & 0.447 & 0.20 & 0.250 & 1.250 & 704 \\\hline
50 & \(\checkmark\) & 100 & 2.236 & 0.000 & 0.20 & 0.250 & 1.250 & 705 \\\hline
51 & \(\checkmark\) & 100 & 1.000 & 0.000 & 0.20 & 0.063 & 0.313 & 707 \\\hline
52 & \(\checkmark\) & 100 & 2.236 & 0.000 & 0.20 & 1.250 & 6.250 & 712 \\\hline
53 & \(\checkmark\) & 100 & 1.673 & 0.000 & 0.20 & 0.700 & 3.500 & 714 \\\hline
54 & \(\checkmark\) & 100 & 0.748 & 0.000 & 0.04 & 0.700 & 3.500 & 715 \\\hline
55 & \(\checkmark\) & 100 & 1.000 & 1.000 & 0.20 & 0.063 & 0.313 & 708 \\\hline
56 & \(\checkmark\) & 400 & 1.000 & 0.000 & 0.20 & 0.250 & 1.250 & 804 \\\hline
\end{tabular} \normalsize
\end{center}
\caption{}
\label{Tbl:Sims}
\end{table}

\addtocounter{table}{-1}
\begin{table}[h]
\caption{Initial inflow parameters for simulations. The column ``\(m_{i}/m_{e}\)
compare''  shows which runs are used in the electron mass ratio comparisons
in Figure~\ref{Fig:massratio}. Values given are ion to electron mass ratio
\((m_{i}/m_{e})\), reconnecting magnetic field strength \((B_{r})\), guide
magnetic field \((B_{g})\), inflowing density \(n_{in}\), inflowing electron
temperature \((T_{e})\), and inflowing ion temperature \((T_{i}). \) The
``Reference Number'' in the final column is for internal indexing of the
runs, and should be used when requesting simulation data from the authors.}
\end{table}

\end{document}